\documentclass{elsart}

\usepackage{amsmath,amssymb,graphicx,psfrag,bm,subfigure}

\newcommand{\im}{\text{Im }}
\newcommand{\Tr}{\text{Tr }}

\newcommand{\cond}[1]{\langle #1 \rangle}

\newcommand{\PDG}{Amsler:2008zzb}

\usepackage[normalem]{ulem}  
\usepackage[dvips]{color} 

\renewcommand\sout{\bgroup \color{red} \ULdepth=-.5ex \ULset}

\begin{document}

\begin{frontmatter}



\title{Nature of the $\sigma$ meson as revealed by its softening process}


\author[TITech]{Tetsuo~Hyodo\corauthref{cor}},
\ead{hyodo@th.phys.titech.ac.jp}
\corauth[cor]{Corresponding author.}
\author[YITP]{Daisuke~Jido},
\author[Kyoto]{Teiji~Kunihiro}

\address[TITech]{Department of Physics, Tokyo Institute of Technology 
 Meguro 152-8551, Japan}
\address[YITP]{Yukawa Institute for Theoretical Physics, 
Kyoto University, \\
Kyoto 606--8502, Japan}
\address[Kyoto]{Department of Physics, Kyoto University, Kyoto 606-8502,
 Japan}

\begin{abstract}
    The $\pi\pi$ scattering is studied in two-flavor chiral models with a finite pion mass to investigate the nature of the $\sigma$ meson which is observed as the lowest scalar-isoscalar resonance. We compare several models with different origins of the $\sigma$ meson, such as the chiral partner of the pion and the dynamically generated $\pi\pi$ molecule. We find that the dynamically generated $\sigma$ meson exhibits a novel pattern of the threshold enhancement reflecting the $s$-wave nature of the resonance, which is qualitatively different from the softening of the chiral partner introduced as a bare field. This behavior around the threshold energy region is universal as far as the bare $\sigma$ pole stays away from the threshold throughout the symmetry restoration process. On the other hand, for $m_{\pi}=0$, the dynamically generated $\sigma$ behaves similarly to the chiral partner in the symmetry restoration limit, implying the possibility of the dynamically generated chiral partner.
\end{abstract}

\end{frontmatter}

\section{Introduction}\label{sec:Intro}

Recent progress in scattering theory~\cite{Caprini:2008fc,Yndurain:2007qm,Caprini:2005zr} as well as the refinement of experimental analyses~\cite{Ablikim:2006bz,Ambrosino:2006hb} have been revealing the precise pole position of the $\sigma$ meson in the $\pi\pi$ scattering amplitude, apart from the pioneering works in Refs.~\cite{Basdevant:1970nu,Elias:1984zh,vanBeveren:1986ea,Tornqvist:1995ay,Harada:1995dc,Ishida:1995xx,Dobado:1997ps,Oller:1997ti,Oller:1998hw,Igi:1998gn} (as a review, see Ref.~\cite{Close:2002zu}). These activities are establishing the existence of the $\sigma$ meson, the lowest scalar-isoscalar resonance in QCD, as it is now listed in the table of the Particle Data Group (PDG)~\cite{\PDG}. Because naive constituent quark models with the $\bar{q}q$ description for the scalar mesons have difficulties in reproducing the light mass of the $\sigma$ and the mass spectrum of scalar nonets, various internal structures beyond the simple $\bar{q}q$ have been proposed: the four-quark state with strong diquark correlation~\cite{Jaffe:1976ig,Black:1998wt,Schafer:2003nu,Ebert:2008id}, the collective $\bar{q}q$ excitation~\cite{Elias:1984zh,Nambu:1961tp,Nambu:1961fr,Hatsuda:1985ey,Hatsuda:1986gu,Hatsuda:1994pi}, the dynamically generated mesonic molecule~\cite{Basdevant:1970nu,Dobado:1997ps,Oller:1997ti,Oller:1998hw} and the glueball~\cite{Minkowski:1998mf}. Although several approaches based on QCD have been performed~\cite{Kojo:2008hk,Alford:2000mm,Kunihiro:2003yj,McNeile:2007fu}, the understanding of the structure of the lowest scalar-isoscalar meson has not been settled yet.

It is our aim to study the internal structure of the $\sigma$ meson. Conventionally, the structure of hadrons has been examined by comparing the prediction of the model with the experimental information such as the mass spectrum, the decay properties, and the scattering phase shifts. An alternative approach is to study the response to the change of the internal/external parameter of the models. For instance, the structure of the $\sigma$ meson was investigated by changing the number of colors ($N_c$) which enables one to extract the $\bar{q}q$ component of the mesons~\cite{Oller:1998zr,Harada:2003em,Pelaez:2003dy,Pelaez:2006nj}. Following the latter philosophy, in this paper, we focus on the softening phenomena of the $\sigma$ meson associated with the partial restoration of chiral symmetry, possibly in hot and/or dense matter.

Although the $\sigma$ meson has a huge decay width in vacuum, it might become a sharp resonance when chiral symmetry is partially restored in a medium with high temperature and/or density~\cite{Hatsuda:1985ey,Hatsuda:1986gu,Hatsuda:1994pi}. In this case the strong enhancement of the $\pi\pi$ cross section is caused near the threshold in the $I=J=0$ channel, which is called the softening of the $\sigma$ meson. There have been some attempts to observe the softening experimentally in order to establish the chiral symmetry restoration, but eventually it turns out that the separation of the medium effect from the final state interaction is difficult (see e.g. \cite{Hayano:2008vn}).

In the linear sigma model~\cite{Hatsuda:1999kd}, the mechanism of the softening is understood as the decrease of the bare mass of the $\sigma$ meson, which leads to the suppression of the phase space and hence the spectrum shows a sharp peak. The threshold enhancement of the cross section is observed also for the dynamically generated $\sigma$ meson without a bare field, owing to the in-medium reduction of the pion decay constant~\cite{Jido:2000bw,Yokokawa:2002pw,FernandezFraile:2007fv}. In this case, the mechanism of the softening would be attributed to the enhancement of the attractive $\pi\pi$ interaction, which changes the resonance into bound state. It is important at this point to recall the property of the dynamically generated $s$-wave resonance. A special nature of the $s$-wave resonance is that, when the attraction is increased, the resonance pole on the second Riemann sheet moves below the threshold keeping finite width, which is called the \textit{virtual state}, prior to becoming the bound state on the first sheet. This will provide a novel softening pattern, which enables us to discriminate the different structure.

Our discussion will be based on a dynamical chiral model in which the $\pi\pi$ scattering amplitude is constructed by respecting the chiral symmetry and unitarity, in order to describe the $\sigma$ meson and the softening phenomena. The $\sigma$ meson is expressed as a resonance pole in the $\pi\pi$ scattering amplitude, and we can classify possible origin of the $\sigma$ meson into three classes: (i) the chiral partner of the pion, (ii) $\pi\pi$ molecule state dynamically generated by the $\pi\pi$ attractive interaction, and (iii) the CDD (Castillejo-Dalitz-Dyson) pole~\cite{Castillejo:1956ed,PR124.264} whose structure originates in the other mechanism than chiral symmetry. Since we study the $\pi\pi$ scattering with the hadronic degrees of freedom, the structures caused by the QCD dynamics, such as the four-quark state and the glueball are classified into this category. Note however that we are not able to pin down the origin of the CDD pole within the present framework. Our classification is similar to the old discussion of the elementarity/compositeness of a particle~\cite{Weinberg:1962hj,Weinberg:1963zz,Weinberg:1965zz}. We examine these structures of the $\sigma$ meson and try to extract the influence of the different structure in the pattern of the softening. Our approach, along the line with Refs.~\cite{Jido:2000bw,Yokokawa:2002pw}, is rather schematic; we utilize two-flavor chiral models and the symmetry restoration is accounted for by the reduction of the chiral condensate $\cond{\sigma}$. It is therefore not our aim to produce a realistic in-medium spectrum for the quantitative comparison with experimental data, for which one may refer to the state-of-the-art calculations~\cite{Roca:2002vd,Cabrera:2005wz,Buss:2006vh,Cabrera:2008tja}. Rather, we want to study the qualitative behavior along with the symmetry restoration, which reflects the structure of the $\sigma$ meson. Although the physical sigma meson should be the mixture of different components, we examine simplified models in which the typical structures of the $\sigma$ meson are realized.

We also consider the properties of the $\sigma$ meson in the symmetry restoration limit for the discussion of the chiral partner of the pion. In Refs.~\cite{Oller:2000wa,Leupold:2008ne} the possibility of the dynamically generated state as the chiral partner was discussed, based on the degeneracy of the masses in the restoration limit. Here we study the coupling strength of the $\sigma$ pole to the $\pi\pi$ scattering state and show that the coupling property is also useful to discriminate the structure of the $\sigma$ meson. 

This paper is organized as follows. We formulate the dynamical chiral models to describe the $\pi \pi$ scattering and the $\sigma$ meson in Section~\ref{sec:vacuum}. The treatment of the symmetry restoration is presented in Section~\ref{sec:restoration}, together with the analysis of the behavior of the scattering amplitude in the symmetry restoration limit. Numerical study is presented in Section~\ref{sec:numerical}; preparing several models with different structure of the $\sigma$ meson, we first analyze the amplitude in vacuum to clarify the origin of the pole, and then introduce the symmetry restoration to study the softening phenomena. The last section is devoted to summary.

\section{The $\pi\pi$ scattering in vacuum}\label{sec:vacuum}

Here we introduce the dynamical models to describe the $\pi\pi$ scattering in $I=0$ and $J=0$ channel. Based on an effective chiral Lagrangian, the tree-level amplitude is derived as the interaction kernel. We extend the amplitude in a way consistent with chiral symmetry by introducing a parameter which governs the strength of the pole contribution. We then unitarize the tree-level interaction through the nonperturbative resummation to obtain the full scattering amplitude.

\subsection{Tree-level amplitude for the $\pi\pi$
scattering}\label{subsec:tree}

To derive the $\pi\pi$ scattering amplitude, we start from the effective chiral Lagrangian of the linear sigma model in which the $\sigma$ and $\pi$ fields are constrained by chiral symmetry and introduced as the chiral partner:
\begin{align}
   \mathcal{L}
   =& \frac{1}{4}\Tr
   \left[
   \partial M \partial M^{\dag} -\mu^2 MM^{\dag}
   -\frac{2\lambda}{4!}(MM^{\dag})^{2}
   +h(M+M^{\dag})
   \right] ,
   \label{eq:Lagrangian} \\
   &M = \sigma + i\bm{\tau}\cdot \bm{\pi}
   \nonumber .
\end{align}
The Lagrangian is invariant under chiral SU(2)$\times$SU(2) symmetry in the limit $h\to 0$. For a negative $\mu^2$, chiral symmetry is spontaneously broken and vacuum expectation value of the $\sigma$ field becomes finite, $\cond{\sigma}\neq 0$. Three parameters in the Lagrangian $\mu$, $\lambda$, and $h$ are determined by the physical quantities in the mean-field level: the chiral condensate $\cond{\sigma}$, the mass of the pion $m_{\pi}$, and the mass of the $\sigma$ meson $m_{\sigma}$. The chiral condensate $\cond{\sigma}$ is the order parameter of the symmetry breaking and coincide with the pion decay constant $f_{\pi}$ at tree level.\footnote{The pion decay constant $f_{\pi}$ is related to the quark condensate in QCD $\cond{\bar{q}q}$ through the Glashow-Weinberg relation $f_{\pi}G_{\pi}^{1/2}=-\cond{\bar{q}q}$ with the coupling constant $G_{\pi}^{1/2}=f_{\pi}m_{\pi}^2/(2m_q)$~\cite{Glashow:1967rx,Jido:2008bk}.}

The Lagrangian~\eqref{eq:Lagrangian} can be rewritten in the nonlinear form by taking $M=(\cond{\sigma}+\tilde{\sigma})U$ and $U=\exp(i\bm{\tau}\cdot\tilde{\bm{\pi}} /\cond{\sigma})$. Since this is nothing but the field redefinition, the nonlinearized Lagrangian gives exactly the same result with the original form with $(\sigma,\bm{\pi})$ fields~\cite{Jido:2000bw}. On the other hand, the nonlinearized $(\tilde{\sigma},\tilde{\bm{\pi}})$ fields are useful to understand the relationship with the chiral perturbation theory where the $\tilde{\sigma}$ degrees of freedom is integrated out. In the following, utilizing both representations, we construct the scattering amplitude in a consistent way with chiral symmetry.

Now let us derive the $\pi\pi$ scattering amplitude at tree level. Crossing symmetry requires the $\pi\pi$ scattering amplitude in general to be
\begin{align}
   T_{\text{tree}}(s,t,u)
   =&
   A(s,t,u)\delta_{ab}\delta_{cd}
   +A(t,s,u)\delta_{ac}\delta_{bd}
   +A(u,t,s)\delta_{ad}\delta_{bc}
   \label{eq:Ttreegeneral} .
\end{align}
From the Lagrangian~\eqref{eq:Lagrangian}, the tree level contribution to the function $A$ is calculated as
\begin{align}
   A(s,t,u)
   \equiv A(s)
   =&-\frac{m_{\sigma}^2-m_{\pi}^2}{\cond{\sigma}^2}
   -\frac{(m_{\sigma}^2-m_{\pi}^2)^2}{\cond{\sigma}^2}
   \frac{1}{s-m_{\sigma}^{2}}
   \nonumber \\
   =&
   \frac{s-m_{\pi}^2}{\cond{\sigma}^2}
   -\frac{(s-m_{\pi}^2)^2}{\cond{\sigma}^2}
   \frac{1}{s-m_{\sigma}^{2}}
   \label{eq:amp} .
\end{align}
The last expression can be also derived from the nonlinearized Lagrangian~\cite{Jido:2000bw}. If the $\sigma$ meson has a large mass and is irrelevant for the low energy scattering, the second term can be dropped and only the first term gives relevant contributions:
\begin{align}
   A_{LO}(s)
   =&
   \frac{s-m_{\pi}^2}{\cond{\sigma}^2} ,
   \label{eq:ampLO}
\end{align}
which corresponds to the leading order contribution to the $\pi\pi$ scattering in the chiral perturbation theory, and thus the second term of Eq.~\eqref{eq:amp} is a part of the next-to-leading order contributions in the low energy expansion. In order to investigate the contribution from the $\sigma$ pole term, we introduce a parameter $x$ in front of the second term as
\begin{align}
   A(s;x)
   =&
   \frac{s-m_{\pi}^2}{\cond{\sigma}^2}
   -x\frac{(s-m_{\pi}^2)^2}{\cond{\sigma}^2}
   \frac{1}{s-m_{\sigma}^{2}} .
   \label{eq:modelext}
\end{align}
This amplitude reduces to Eq.~\eqref{eq:amp} with $x=1$ and to Eq.~\eqref{eq:ampLO} with $x=0$. It is worth noting that both Eq.~\eqref{eq:amp} and Eq.~\eqref{eq:ampLO} [and therefore the second term of Eq.~\eqref{eq:amp}] satisfy the Adler condition $A(m_{\pi}^2,m_{\pi}^2,m_{\pi}^2) =0$~\cite{Adler:1964um}. This means that the amplitude~\eqref{eq:modelext} is consistent with chiral low energy theorem, even away from $x=1$ and $x=0$; indeed the consistency is clearly seen by the decomposition $A(s;x)=xA(s)+(1-x)A_{LO}(s)$. The amplitude used in Ref.~\cite{Igi:1998gn} (and \textit{model B} in Ref.~\cite{Yokokawa:2002pw}) which is motivated by degeneracy of the $\rho$ and $\sigma$ meson and KSRF relation can be obtained by taking $x=1/2$ and $m_{\pi}\to 0$ in Eq.~\eqref{eq:modelext}.\footnote{There is a difference from Ref.~\cite{Yokokawa:2002pw} in the treatment of the $\cond{\sigma}$-dependence of the coupling constant. We will come back to this point later in the analysis of the chiral symmetry restoration.} As we will see below, extrapolation of the models by varying the parameter $x$ is useful to study the origin of the resonance. 

To study the $\sigma$ meson, we project the amplitude onto the $I=J=0$ channel in the center-of-mass frame
\begin{align}
    T_{\text{tree}}(s;x)
    =& \frac{1}{2}\int_{-1}^{1}d \cos\theta [3A(s;x)+A(t;x)+A(u;x)]
    \nonumber \\
    =& \frac{m_{\sigma}^{2}-m_{\pi}^2}{\cond{\sigma}^2}
    \Biggl[
    \frac{2s-m_{\pi}^2}{m_{\sigma}^{2}-m_{\pi}^2}(1-x)-5x 
    \nonumber \\
    &-3x\frac{m_{\sigma}^{2}-m_{\pi}^2}{s-m_{\sigma}^{2}}
    -2x\frac{m_{\sigma}^{2}-m_{\pi}^2}{s-4m_{\pi}^2}
    \ln\left(\frac{m_{\sigma}^{2}}{m_{\sigma}^{2}+s-4m_{\pi}^2}\right)
    \Biggr] ,
    \label{eq:Ttree}
\end{align}
where we have used the relation in the center-of-mass frame
\begin{equation}
    t= -
    \left(\frac{s}{2}-2m_{\pi}^2\right)
    (1-\cos\theta), \quad
    u=-
    \left(\frac{s}{2}-2m_{\pi}^2\right)(1+\cos\theta).
    \nonumber
\end{equation}
Clearly, the familiar results of the linear sigma model~\cite{Basdevant:1970nu,Achasov:1994iu} and chiral perturbation theory are reproduced by $x=1$ and $x=0$, respectively,
\begin{align}
    T_{\text{tree}}(s;1)
    =& -\frac{m_{\sigma}^{2}-m_{\pi}^2}{\cond{\sigma}^2}
    \Biggl[
    5+3\frac{m_{\sigma}^{2}-m_{\pi}^2}{s-m_{\sigma}^{2}}
    +2
    \frac{m_{\sigma}^{2}-m_{\pi}^2}{s-4m_{\pi}^2}
    \ln
    \frac{m_{\sigma}^{2}}{m_{\sigma}^{2}+s-4m_{\pi}^2}
    \Biggr]
    \label{eq:TtreeL} , \\
    T_{\text{tree}}(s;0)
    =&\frac{1}{\cond{\sigma}^2}
    [2s-m_{\pi}^2]
    \label{eq:TtreeNL} .
\end{align}
In the present work, we concentrate on the property of the $\sigma$ meson and we do not include the effect of the $\rho$ meson in $I=J=1$ channel. It is known that the standard linear sigma model does not provide the empirical low energy constants of the chiral perturbation theory~\cite{Gasser:1983yg} and the next-to-leading order terms are saturated by the $\rho$ meson pole contribution~\cite{Ecker:1988te}, so the $\rho$ should be included for a quantitative calculation of the $\pi\pi$ scattering amplitude. However, in the present study, we deal with the softening phenomena of the $\sigma$ meson in $I=J=0$ channel, where the effect of the $\rho$ exchange does not directly contribute as the $s$-channel diagram, as far as we consider the isospin symmetric environment. Moreover, the low energy amplitude around threshold should be governed by the leading order interaction, while the $t$-channel $\rho$ exchange contributes to the next-to-leading order terms. Indeed, as studied in Ref.~\cite{Yokokawa:2002pw}, the existence of the $\rho$ meson exchange term does not alter the qualitative feature of the softening of the $\sigma$ meson near the restoration limit, since the $\sigma$ pole approaches the low energy region where the amplitude is dominated by the leading order term.

\subsection{Sign of the contact interaction}\label{subsec:contact}

For later convenience, we study the sign of the interaction~\eqref{eq:Ttree} for different values of $x$. We divide the tree-level amplitude~\eqref{eq:Ttree} into the ``pole'' term and the ``contact'' term, by isolating the $\sigma$ pole contribution:
\begin{align}
    T_{\text{tree}}(s;x)
    \equiv &  T_{\text{tree}}^{\text{(contact)}}(s;x) 
    + T_{\text{tree}}^{\text{(pole)}}(s;x)
    \nonumber , \\
    T_{\text{tree}}^{\text{(pole)}}(s;x)
    =&-3x \frac{(m_{\sigma}^{2}-m_{\pi}^2)^2}{\cond{\sigma}^2}
    \frac{1}{s-m_{\sigma}^{2}} \nonumber , \\
    T_{\text{tree}}^{\text{(contact)}}(s;x) 
    =& \frac{m_{\sigma}^{2}-m_{\pi}^2}{\cond{\sigma}^2}
    \Biggl[
    \frac{2s-m_{\pi}^2}{m_{\sigma}^{2}-m_{\pi}^2}(1-x) \nonumber \\
    &-5x -2x\frac{m_{\sigma}^{2}-m_{\pi}^2}{s-4m_{\pi}^2}
    \ln\left(\frac{m_{\sigma}^{2}}{m_{\sigma}^{2}+s-4m_{\pi}^2}\right)
    \Biggr]
    \nonumber .
\end{align}
The pole term is chosen such that the residue is energy independent, and the contact term is defined as those which are not singular at the pole position as functions of the energy. In this definition, the contribution from the pole term is dominant only at the energy region around the pole mass, while the other energy region is dominated by the contact term. 

In order to investigate the sign of the contact interaction, we consider 
asymptotic behavior of $T_{\text{tree}}^{\text{(contact)}}(s;x) $ at $s\to \infty$:
\begin{align}
    \left.
    T_{\text{tree}}^{\text{(contact)}}(s;x) \right|_{s\to \infty}
    =& 
    \begin{cases}
        +\infty & x<1  \\
        -5\frac{m_{\sigma}^2-m_{\pi}^2}{\cond{\sigma}^2} & x=1  \\
        -\infty & x>1 
    \end{cases} .
    \label{eq:Tinfty}
\end{align}
At the threshold $(s=4m_{\pi}^2)$, the amplitude behaves as
\begin{equation}
    \left.T_{\text{tree}}^{\text{(contact)}}(s;x) \right|_{s\to 4m_{\pi}^2}
    \begin{cases}
        >0 & x<C  \\
        =0 & x=C  \\
        <0 & x>C
    \end{cases} ,
    \label{eq:Tth}
\end{equation}
with the critical value of $x$ being 
\begin{equation}
    C=\frac{7}{3 m_{\sigma}^{2}/m_{\pi}^2
    +6-2m_{\pi}^2/m_{\sigma}^{2}}
    \label{eq:const} ,
\end{equation}
where $0<C<1$ for $m_{\sigma}>m_{\pi}>0$. In the chiral limit ($m_{\pi}^2=0$), $C=0$. For $m_{\pi}=140$ MeV and $m_{\sigma}=550$ MeV, we obtain $C\sim 0.134$.

Combining Eqs.~\eqref{eq:Tinfty} and \eqref{eq:Tth}, we summarize the sign of the contact interaction in Table~\ref{tbl:contact}. For $x<C$ ($x\geq 1$) the interaction is attractive (repulsive) for whole energy region, while the sign of the amplitude depends on the energy for $C<x<1$. We find that the contact interaction for $x<1$ is attractive at least for some energy region above the threshold. The existence of an attraction is crucial for the dynamical generation of resonance, as we will see below.

\begin{table}[tbp]
    \centering
    \caption{Sign of the contact interaction at $s\to \infty$ and at the threshold with parameter $x$. The constant $C$ is given in Eq.~\eqref{eq:const}.}
    \begin{tabular}{c|ccccc}
    \hline
     & $x<C$ & $x=C$ & $C<x<1$ & $x=1$ & $x>1$  \\
    \hline
    $T_{\text{tree}}^{\text{(contact)}}(s;x) |_{s\to \infty}$
    & $+$ & $+$ & $+$ & $-$ & $-$  \\
    $T_{\text{tree}}^{\text{(contact)}}(s;x) |_{s\to 4m_{\pi}^2}$
    & $+$ & $0$ & $-$ & $-$ & $-$  \\
    \hline
    \end{tabular}
    \label{tbl:contact}
\end{table}

Eq.~\eqref{eq:Tth} shows that the contact interaction at the threshold changes the sign, depending on the parameter $x$. This sounds contradicting with chiral low energy theorem for the $\pi\pi$ scattering length~\cite{Weinberg:1966kf}. Note however that the $\pi\pi$ scattering length is given by the sum of the $T_{\text{tree}}^{\text{(contact)}}(4m_{\pi}^2,x)$ and $T_{\text{tree}}^{\text{(pole)}}(4m_{\pi}^2,x)$, namely there is another contribution from the pole term. Indeed, in the form of Eq.~(6), the leading order contribution in the low energy expansion is consistent with the chiral theorem, irrespective of the value of $x$. The effect of the (higher order) $\sigma$ pole term gives a deviation of the scattering length from the value of the low energy theorem. In the numerical analysis (Table~\ref{tbl:modeldata}), we show the deviation of the scattering lengths for several values of the parameter $x$.

\subsection{Unitarization of the amplitude}\label{subsec:unitarization}

The tree-level amplitude increases with energy so that it violates the unitarity at a certain kinematical scale. From the optical theorem, the unitarity condition is given by
\begin{equation}
    \im T^{-1}(s) = -\frac{\Theta(s)}{2}
    \quad \text{for}\quad s>4m_{\pi}^2 ,
    \nonumber
\end{equation}
where $\Theta(s)=(16\pi)^{-1}\sqrt{1-4m_{\pi}^2/s}$ is the two-body phase space function. A simple way to obtain the unitarized amplitude is the inverse amplitude method used for instance in Refs.~\cite{Achasov:1994iu,Jido:2000bw}. Here we utilize the prescription in Ref.~\cite{Oller:1998zr} based on the N/D method, where the real part of the amplitude is determined by the imaginary part to satisfy the dispersion relation, so the analyticity is preserved. In this method, we decompose the scattering amplitude $T(s)$ into the numerator $N(s)$ and the denominator $D(s)$ which are responsible for the unphysical cut and the unitarity cut, respectively:
\begin{align}
    T(s)
    =&\frac{N(s)}{D(s)} , \nonumber \\
    D(s)
    =& -\frac{1}{2\pi}
    \int_{s_{\text{th}}}^{\infty}
    ds^{\prime}
    \frac{\Theta(s^{\prime})N(s^{\prime})}{s^{\prime}-s}
    +(\text{subtractions})+(\text{pole terms}) ,
    \nonumber \\
    N(s)
    =& \frac{1}{2\pi}
    \int_{-\infty}^{s_{\text{left}}}
    ds^{\prime}
    \frac{2\im T(s^{\prime})D(s^{\prime})}
    {s^{\prime}-s}
    +(\text{subtractions})+(\text{pole terms})
    \label{eq:left} ,
\end{align}
where $s_{\text{th}}=4m_{\pi}^2$ and $s_{\text{left}}=0$ for the $\pi\pi$ scattering and the subtractions are introduced to make the integration convergent and possible CDD poles can contribute to the dispersion relations~\cite{Oller:1998zr}. The left hand cut is responsible for the crossed channel dynamics. Since the left hand cut lies in the subthreshold region, its effect to the resonances above the threshold is in general considered to be small. Neglecting the contribution from the left hand cut~\eqref{eq:left} and putting $N(s)=1$, we obtain a general expression of the scattering amplitude which satisfies the unitarity condition. Matching the chiral interaction $T_{\text{tree}}(s;x)$ in the loop expansion of the full amplitude $T(s;x)$, the unitarized amplitude which is consistent with chiral low energy theorem is given as
\begin{align}
    T(s;x)
    =& \frac{1}{T_{\text{tree}}^{-1}(s;x)+G(s)} 
    \label{eq:TChU} , \\
    G(s)
    =& \frac{i}{2}\int\frac{d^{4}q}{(2\pi)^{4}}
    \frac{1}{(P-q)^{2}-m_{\pi}^{2}+i\epsilon}
    \frac{1}{q^{2}-m_{\pi}^{2}+i\epsilon}
    \nonumber \\
    =& \frac{1}{2}\frac{1}{(4\pi)^2}
    \Biggl\{
    a(\mu)+\ln\frac{m_{\pi}^2}{\mu^2} \\
    &+\sqrt{1-\frac{4m_{\pi}^2}{s}}
    \left[\ln
    \left(
    \sqrt{1-\frac{4m_{\pi}^2}{s}}+1\right)
    -\ln
    \left(\sqrt{1-\frac{4m_{\pi}^2}{s}}-1\right)
    \right]
    \Biggr\} \nonumber ,
\end{align}
where $a(\mu)$ is the subtraction constant at the subtraction scale $\mu$. Because of the rescaling relation $a(\mu^{\prime})=a(\mu)+2\ln(\mu^{\prime}/\mu)$, there is one degree of freedom of cutoff, which corresponds to the single subtraction in the dispersion theory. In the following we choose the subtraction scale at $\mu=m_{\pi}$ and denote the subtraction constant at this scale as $a\equiv a(m_{\pi})$. 

Let us consider the pole singularities of the full amplitude $T$. 
If the tree-level amplitude has the pole of the bare $\sigma$ state on the real axis (e.g. $x\neq 0$ case), then the bare state acquires a finite width in the full amplitude, through the coupling to the $\pi\pi$ state. On the other hand, even if the tree-level interaction does not contain the bare state, a sufficiently strong attractive interaction will generate a resonance dynamically in the full amplitude, since Eq.~\eqref{eq:TChU} corresponds to the infinite resummation.  

Since the amplitude has the unitarity cut on the real axis above the threshold, the complex energy plane has two Riemann sheets. Causality prohibits the existence of the pole singularity in the first Riemann sheet except for the bound states, so the pole corresponding to the resonance state should appear in the second Riemann sheet. In the expression~\eqref{eq:TChU}, the analytic structure of the amplitude $T(s;x)$ is determined by the loop function $G(s)$, and therefore the amplitude in the second Riemann sheet $T_{II}$ is given by
\begin{align}
    T_{II}(s;x)
    =& \frac{1}{T_{\text{tree}}^{-1}(s;x)+G_{II}(s)} 
    \nonumber , \\
    G_{II}(s)
    =&G(s)
    +\frac{1}{2}\frac{2\pi i}{(4\pi)^2}
    \sqrt{1-\frac{4m_{\pi}^2}{s}}\nonumber .
\end{align}
We use this expression when we search for the resonance pole. 

As we mentioned, the bound state pole appears on the real axis in the first 
Riemann sheet below the threshold, and the resonance pole appears in the complex energy plane in the second Riemann sheet above the threshold. In the $s$-wave scattering case, there is another class of singularity, called the virtual state, as is known for the spin-singlet deuteron~\cite{Bohm:2001,BlattWeisskopf}. Theoretically, the virtual state is expressed by the \textit{pole in the second Riemann sheet below the threshold}. The virtual state pole may be accompanied by the finite imaginary part, even below the threshold. Experimentally, we can observe its remnant as an enhancement of the spectrum near the threshold together with a large attractive scattering length. We shall see that the virtual state plays an important role in the softening of the dynamically generated $\sigma$ meson.

In the loop function $G(s)$, the subtraction constant $a$ should in principle be determined by fitting experimental data, in order to compensate the effects which are not included in the model setup. However, to concentrate on the dynamical nature of the resonances, here we determine the subtraction constant by excluding the nontrivial CDD pole in the amplitude~\cite{Hyodo:2008xr}. This can be achieved by imposing the condition
\begin{equation}
    G(s)=0 \quad \text{at} \quad
    s=m_{\pi}^2 
    \label{eq:Gcond},
\end{equation}
which leads to
\begin{equation}
    a
    =-\frac{\pi}{\sqrt{3}}
    \label{eq:ampi}.
\end{equation}
With the subtraction constant~\eqref{eq:ampi}, the scattering amplitude $T$ 
reduces into the tree level one $T_{\text{tree}}$ at $s=m_{\pi}^2$. This 
condition will be important when we discuss the properties of the $\sigma$ meson in the restoration limit. Similar conditions have been used in a different context, for instance, through the consistency with the amplitude of chiral perturbation theory~\cite{Meissner:1999vr,Lutz:2001yb} and through the matching with the $u$-channel amplitude~\cite{Igi:1998gn,Lutz:2001yb}. 

In the present context, we emphasize that Eq.~\eqref{eq:Gcond} can be used to single out the origin of the resonance~\cite{Hyodo:2008xr}. It has been shown in Ref.~\cite{Hyodo:2008xr} that the condition~\eqref{eq:Gcond} excludes the possible CDD pole in the loop function. Therefore, under the renormalization condition~\eqref{eq:Gcond}, the origin of the resonance except for the dynamically generated one can be attributed to the pole term in the interaction kernel $T_{\text{tree}}$.

The present unitarization method is based on Ref.~\cite{Oller:1998zr}, where we put $N=1$ in the derivation of the amplitude. The framework used in Refs.~\cite{Igi:1998gn,Yokokawa:2002pw}, on the other hand, sets $N=T_{\text{tree}}$ (single N/D iteration). For more quantitative analysis of the $\pi\pi$ scattering, one may also take into account the effect of the left hand cut. However, we numerically checked that our model is qualitatively consistent with Ref.~\cite{Yokokawa:2002pw} in the limit $m_{\pi}\to 0$, so we expect that the difference of the N/D framework would not change the qualitative feature of the softening drastically, and that the amplitude~\eqref{eq:TChU} is sufficient for the present purpose. 

\section{Chiral symmetry restoration}\label{sec:restoration}

Here we consider the restoration of chiral symmetry and discuss the chiral partner of the pion. Let us first consider the property of the chiral partner based on symmetry principle. Lagrangian~\eqref{eq:Lagrangian} clearly shows that the mass of the $\sigma$ is degenerated with the pion mass in the Wigner phase where $\cond{\sigma}= 0$. It is also observed that there is no three-point vertex for $\cond{\sigma}= 0$, which indicates that the $\sigma\pi\pi$ coupling constant should vanish in the restoration limit. Thus, we adopt the conditions for the $\sigma$ meson as the chiral partner of the pion in the restoration limit as
\begin{itemize}
    \item[(i)]  the degeneracy of the mass with the pion and

    \item [(ii)] vanishing of the coupling to the $\pi\pi$ scattering state.
\end{itemize}
In the following, we first introduce the effect of the symmetry restoration in the present model, and then analyze the properties of the $\pi\pi$ scattering amplitude in the restoration limit.

\subsection{Prescription for chiral symmetry restoration}\label{subsec:model}

In this study, we introduce the effect of the symmetry restoration from the outside of the model, by changing the parameter of the model. Chiral condensate should decrease with the chiral symmetry restoration, so we parametrize the condensate by
\begin{equation}
    \cond{\sigma}= \Phi \cond{\sigma}_0,
    \quad 0\leq \Phi\leq 1 ,
    \label{eq:condensate}
\end{equation}
where $\cond{\sigma}_0$ is the condensate in vacuum and $\Phi$ is the parameter which conducts the symmetry restoration; $\Phi=1$ corresponds to the vacuum without symmetry restoration and $\Phi=0$ to the restoration limit. This treatment may be justified by the mean-field contribution to the in-medium modification of the condensate in the linear sigma model. It is also the case for the nonlinear Lagrangian with proper renormalization of the pion field~\cite{Jido:2000bw,Yokokawa:2002pw}. 

There are two more parameters in the model, $m_{\pi}$ and $m_{\sigma}$ whose dependence on the symmetry restoration should be also specified. The behavior of the pion mass at finite temperature/density has been studied in various approaches. The weak dependence of the pion mass on the symmetry restoration, at least for low temperature/density, has been found in the linear sigma model~\cite{Larsen:1985ei,Ayala:2000px}, the Landau mean-field theory~\cite{Barducci:1991rh}, the NJL model~\cite{Hatsuda:1985ey,Hatsuda:1986gu,Hatsuda:1994pi}, the chiral perturbation theory~\cite{Schenk:1993ru}, the QCD sum rule~\cite{Kodama:1995kj}, and the Dyson Schwinger equation model~\cite{Bender:1996bm}. We therefore assume that the mass of the pion does not change:
\begin{align}
    m_{\pi}
    = \text{const.}
    \nonumber
\end{align}
For the amplitude without bare $\sigma$ pole ($x=0$), this completes the prescription for the symmetry restoration. For $x\neq 0$, we need to specify the property of the bare $\sigma$ mass $m_{\sigma}$.

If the $\sigma$ meson is the chiral partner of the pion, the mass of the bare $\sigma$ should be degenerated with the pion when the symmetry is restored:
\begin{align}
    m_{\sigma}\bigl|_{\cond{\sigma}\to 0} 
    = m_{\pi} \quad \text{(case I)},
    \label{eq:case1}
\end{align}
which can be achieved by
\begin{align}
    m_{\sigma}
    =\sqrt{\lambda \frac{\cond{\sigma}^2}{3}+m_{\pi}^2} ,
    \nonumber
\end{align}
with $\lambda$ and $m_{\pi}$ being fixed. This is similar to the treatment of Ref.~\cite{Jido:2000bw}. On the other hand, we may consider that the bare $\sigma$ has different origin from chiral symmetry (the CDD pole). In this case, we assume that the mass of the bare $\sigma$ should be unchanged:
\begin{align}
    m_{\sigma}
    = \text{const.} \quad \text{(case II)}.
    \label{eq:case2}
\end{align}
This is similar prescription with Ref.~\cite{Yokokawa:2002pw}. 

In this study we do not introduce the medium effect to the loop function $G(s)$. One may consider the medium modification of the pion propagator at finite temperature~\cite{FernandezFraile:2007fv} or density~\cite{Cabrera:2005wz,Cabrera:2008tja}, as well as the scale dependence of the cutoff value which results in the modification of the renormalization constant~\cite{Geng:2008ag}. Our strategy here is to prepare the purified model in vacuum, and extrapolate it to the symmetry restored world by changing the interaction kernel. We keep the renormalization condition in vacuum throughout the symmetry restoration process, and let the change of $\cond{\sigma}$ in the interaction kernel be responsible for the chiral symmetry restoration.

Let us summarize the possible $\cond{\sigma}$-dependence of $m_{\sigma}$ for different values of parameter $x$. For $x=1$, both case I and case II are allowed in principle. The sigma field is absent for $x=0$, so the property of the $\sigma$ mass is irrelevant in this case. For a model with $x\neq 1$, we adopt case II, so that the origin of the $\sigma$ is attributed to the CDD pole.

\subsection{Behavior of the $\sigma$ in the restoration limit}
\label{subsec:limit}

Without the symmetry restoration, the full scattering amplitude has a resonance pole of the sigma meson. Here we would like to study the fate of the resonance in the restoration limit, by looking at the behavior of the $\pi\pi$ scattering amplitude with $\cond{\sigma}$ being decreased to zero. We compare two cases: the amplitude with $x=1$ and the case I for the $\cond{\sigma}$-dependence of $m_{\sigma}$, and the case II for the $m_{\sigma}$ with an arbitrary $x$. The former corresponds to the sigma meson as the chiral partner of the $\pi$, while the latter to the dynamically generated sigma ($x=0$) or to the CDD pole contribution ($x\neq 0$).

We first consider the case I with $x=1$. We rewrite the tree level amplitude to visualize the $\cond{\sigma}$-dependence in $m_{\sigma}$ as
\begin{align*}
    T_{\text{tree}}(s;1)
    =&
    -\frac{5\lambda}{3}
    -\frac{\lambda^2\cond{\sigma}^2}{3}
    \frac{1}{s-m_{\pi}^2-\frac{\lambda}{3}\cond{\sigma}^2} \\
    &-\frac{2\lambda^2\cond{\sigma}^2}{9}
    \frac{1}{s-4m_{\pi}^2}
    \ln \frac{m_{\pi}^2+\frac{\lambda}{3}\cond{\sigma}^2}
    {s-3m_{\pi}^2+\frac{\lambda}{3}\cond{\sigma}^2}
    \label{eq:LSMamp} .
\end{align*}
In this expression, chiral symmetry restoration is achieved by taking $\cond{\sigma} \to 0$ with $\lambda$ and $m_{\pi}$ being fixed. The second term represents the bare pole of the $\sigma$ meson. As $\cond{\sigma}\to 0$, the mass of the bare $\sigma$ decreases and finally it coincides with the pion mass. Because our renormalization condition requires $G(s)=0$ at $s=m_{\pi}^2$, the full amplitude $T(s;1)$ reduces to the tree level one $T_{\text{tree}}(s;1)$ at $s=m_{\pi}^2$ [see Eq.~\eqref{eq:Gcond}]. So the full amplitude $T(s;1)$ also has a pole at the pion mass in the restoration limit. To extract the mass of the state $M_{\text{pole}}$ and the coupling to the scattering state $g$, we approximate the amplitude by the Breit-Wigner form around the pole as
\begin{equation}
    T(s;1)\sim -\frac{g^2}{s-M^2_{\text{pole}}} .
    \label{eq:BW}
\end{equation}
In the present case, we find
\begin{equation}
    g\sim \frac{\lambda\cond{\sigma}}{\sqrt{3}} ,
    \quad M_{\text{pole}}\sim
    \sqrt{m_{\pi}^2+\frac{\lambda}{3}\cond{\sigma}^2}
    \nonumber ,
\end{equation}
so that 
\begin{equation}
    g\to 0 ,\quad M_{\text{pole}}\to m_{\pi}
    \quad \text{for}\quad \cond{\sigma}\to 0
    \quad (\text{case I})
    \label{eq:limitcaseI} .
\end{equation}
Namely, as anticipated, the mass of the $\sigma$ meson is degenerated with the pion mass, and the coupling to the $\pi\pi$ scattering state vanishes in the symmetry restoration limit. This satisfies the conditions for the chiral partner (i) and (ii) we mentioned above.

Next, we discuss the case II where $m_{\sigma}$ is independent of $\cond{\sigma}$. Strictly speaking, in the nonlinear realization, theory is not defined at $\cond{\sigma} = 0$, but we can investigate the asymptotic form of the amplitude when we approach the restoration limit. For the case II, $\cond{\sigma}$-dependence of the tree-level amplitude~\eqref{eq:Ttree} stems from the overall factor,
\begin{equation}
    T_{\text{tree}}(s;x)
    \propto \frac{1}{\cond{\sigma}^2} .
\end{equation}
Taking the restoration limit $\cond{\sigma}\to 0$, this term diverges, and therefore the full amplitude is solely determined by the loop function $G(s)$, irrespective to the value of $x$ and $m_{\sigma}$:
\begin{equation}
    T(s;x)=\frac{1}{T_{\text{tree}}^{-1}(s,x)+G(s)}
    \to \frac{1}{G(s)}\quad \text{for}\quad \cond{\sigma}\to 0 .
\end{equation}
Thus, for case II, the pole of the amplitude is given by the zero of $G(s)$ in the restoration limit. Interestingly, with the present renormalization scheme of Eqs.~\eqref{eq:Gcond} and \eqref{eq:ampi}, we require $G(s)=0$ for $s=m_{\pi}^2$. This means that the renormalization scheme guarantees the existence of a pole at the pion mass in the $\sigma$ channel.

Since the analytic form of the $G(s)$ function is known, the coupling constant $g$ can be evaluated by calculating the residue of this pole:
\begin{align}
    g^2|_{\cond{\sigma}\to 0}
    =&-(s-m_{\pi}^2)T(s)\bigr|_{s\to m_{\pi}^2,\cond{\sigma}\to 0}\nonumber \\
    =&\left.-\frac{s-m_{\pi}^2}{G(s)}\right|_{s\to m_{\pi}^2}\nonumber \\
    =&(4\pi)^2
    \left(\frac{\pi}{3\sqrt{3}}-\frac{1}{2}\right)^{-1}m_{\pi}^2 .
    \label{eq:NLScoupling}
\end{align}
Thus we obtain a finite coupling constant which is proportional to $m_{\pi}$ and positive definite. This may be a reasonable result, since the coupling constant $g$ has the dimension of mass in the present definition, and in the restoration limit ($\cond{\sigma}\to 0$) the pion mass is the only quantity which has mass dimension. Therefore, the coupling constant should be proportional to $m_{\pi}$, if it does not vanish. In summary, we obtain
\begin{equation}
    g\to 4\pi
    \left(\frac{\pi}{3\sqrt{3}}-\frac{1}{2}\right)^{-1/2}m_{\pi} 
     ,\quad M_{\text{pole}}\to m_{\pi}
    \quad \text{for}\quad \cond{\sigma}\to 0
    \quad (\text{case II})
    \label{eq:limitcaseII} .
\end{equation}

Let us consider the implication of this result to the chiral partner. Strictly speaking, the notion of the chiral partner is defined only in the chiral limit ($m_{\pi}\to 0$), where the SU(2)$\times$SU(2) symmetry is exact in the Wigner phase. In the chiral limit, both Eqs.~\eqref{eq:limitcaseI} and \eqref{eq:limitcaseII} indicate
\begin{equation}
    g\to 0 
     ,\quad M_{\text{pole}}\to 0
    \quad \text{for}\quad \cond{\sigma}\to 0
    \quad (\text{chiral limit})
    \label{eq:limitchirallimit} .
\end{equation}
so the asymptotic value of the mass and coupling constant of the $\sigma$ pole is exactly the same with each other. Note that for $x=0$, there is no bare $\sigma$ state in the beginning, so the pole of the amplitude corresponds to the dynamically generated $\sigma$ meson. Eq.~\eqref{eq:limitchirallimit} shows that the amplitude in this case also has a pole which behaves like the chiral partner in the restoration limit. This implies that the dynamically generated $\sigma$ meson behaves as the chiral partner of the pion in the chiral limit. It should be mentioned that in the present calculation we have neglected the left hand cut, whose effect will become important when we decrease the pion mass, since the threshold energy $s_{\text{th}}=4m_{\pi}^2$  comes closer to the branch point of the left hand cut, $s_{\text{left}}=0$. 

The possibility that the dynamically generated $\sigma$ can be the chiral partner has been discussed in Refs.~\cite{Oller:2000wa,Leupold:2008ne} based on the mass degeneracy in the chiral limit. Here we study the property of the $\sigma$ pole with explicit symmetry breaking, and evaluate not only the asymptotic value of the mass but also the coupling constant to the $\pi\pi$ scattering state. This analysis shed new light on the scenario of dynamically generated $\sigma$ meson as the chiral partner. In the present model, the renormalization condition~\eqref{eq:Gcond}, which was introduced through the matching with the chiral low energy theorem~\cite{Hyodo:2008xr,Meissner:1999vr,Lutz:2001yb}, guarantees the mass degeneracy of the $\sigma$ and $\pi$.

\section{Numerical analysis}\label{sec:numerical}

\subsection{Structure of the $\sigma$ meson in vacuum}\label{subsec:vacuum}

Here we perform numerical calculation to study the property of the $\sigma$ meson with the symmetry restoration. The parameters in the Lagrangian are fixed by demanding $\cond{\sigma}_0=93$ MeV, $m_{\pi}=140$ MeV, and $m_{\sigma}=550$ MeV in vacuum. We use the parameter $x$ and the bare $\sigma$ mass behavior with the symmetry restoration (case I and II) to characterize possible structure for the $\sigma$ meson. 

We first consider the case where the $\sigma$ meson is the chiral partner of the pion as in the linear sigma model. This can be realized by taking $x=1$ and case I, to which we refer as ``model A''. If the $\sigma$ meson is not the chiral partner, there are two possibilities for the origin: dynamical state generated by the $\pi\pi$ interaction or the CDD pole created by mechanisms other than chiral dynamics. To construct the purely dynamical state, we choose $x=0$ and call it ``model B''. In this model, there is no bare $\sigma$ propagator in the interaction and the present renormalization condition excludes the CDD pole in the loop function. Since there is no bare $\sigma$ meson, the prescription of $M_{\sigma}$ for the symmetry restoration is not relevant here.

When the $\sigma$ meson is accounted for by the CDD pole, its physical origin would be, for instance, the four-quark state with strong diquark correlation, the glueball, or some more exotic structures including their mixings. We expect that the bare mass of the $\sigma$ should not strongly depend on the symmetry restoration, so we adopt case II for the $\cond{\sigma}$-dependence of the $m_{\sigma}$. We further classify this case by the sign of the contact interaction as discussed in Sec.~\ref{subsec:contact}. Taking $x=1$, the contact interaction is always repulsive (model C). With $x=1/2$, the contact interaction is attractive for higher energy region (model D). In vacuum, model C is identical to model A. In this way, we prepare altogether four models as summarized in Table~\ref{tbl:model}. 

\begin{table}[tbp]
    \centering
    \caption{Properties of the models: value of the parameter $x$,
    sign of the contact term,
    $\cond{\sigma}$-dependence of $m_{\sigma}$, 
    number of poles in the full amplitude $T$, 
    and possible origin of the pole(s).}
    \begin{tabular}{l|ccccl}
    \hline
     & $x$ & contact term & $m_{\sigma}$ & num. of poles & origin\\
    \hline
    model A & 1 & repulsive  & case I & 1 & CDD (chiral partner) \\
    model B & 0 & attractive & (irrelevant) & 1 & dynamically generated  \\
    model C & 1 & repulsive  & case II & 1 & CDD  \\
    model D  & 1/2 & partly attractive & case II & 2 & CDD + dynamically generated\\
    \hline
    \end{tabular}
    \label{tbl:model}
\end{table}

\begin{table}[tbp]
    \centering
    \caption{
    The scattering length~\eqref{eq:length} by the tree level 
    ($a_{\text{tree}}$) and by the full amplitude ($a$) in units of 
    $m_{\pi}^{-1}$, and pole positions of the amplitude in vacuum. 
    Empirical values for these quantities are 
    $a_{\text{exp}}\sim 0.222m_{\pi}^{-1}$~\cite{Batley:2007zz,BlochDevaux:2009zzb} and $z=441-272i$ 
    MeV~\cite{Caprini:2005zr}, respectively.}
    \begin{tabular}{l|lll}
    \hline
     & $a_{\text{tree}}$ [$m_{\pi}^{-1}$] & $a$ [$m_{\pi}^{-1}$] & pole positions in vacuum [MeV] \\
    \hline
    model A, C & 0.214 & 0.244 & $423-126 i$  \\
    model B & 0.158 & 0.174 & $364-356 i$ \\
    model D & 0.186 & 0.208 & $512-162 i$, \ \ $732-295 i$ \\
    \hline
    \end{tabular}
    \label{tbl:modeldata}
\end{table}

Let us show the vacuum properties of the models. We first calculate the $I=0$ $\pi\pi$ scattering length $a$ (in units of $m_{\pi}^{-1}$):
\begin{equation}
    a
    =\frac{1}{32\pi}T(s=4m_{\pi}^2)
    \label{eq:length} .
\end{equation}
In Table~\ref{tbl:modeldata} we show the result of $a$ calculated by the tree level amplitude $T_{\text{tree}}$ and the full amplitude $T$ in each model. The result can be compared with the Weinberg's low energy theorem $a_W$~\cite{Weinberg:1966kf,Gasser:1983kx} and the central value of the experimental determination $a_{\text{exp}}$ from $K_{e4}$ decay~\cite{Batley:2007zz,BlochDevaux:2009zzb}\footnote{Here we adopt the value of the scattering length extracted from the analysis of the $K_{e4}$ decay with isospin breaking correction~\cite{Colangelo:2008sm} as reported in Ref.~\cite{BlochDevaux:2009zzb}.}:
\begin{equation}
    a_{W}
    =\frac{7m_{\pi}^2}{32\pi\cond{\sigma}_0^2}
    \sim
    0.158 m_{\pi}^{-1},
    \quad 
    a_{\text{exp}}
    \sim 0.222 m_{\pi}^{-1}.
    \label{eq:lengthex}
\end{equation}
Tree level result of model B corresponds to the value of the low energy theorem $a_{W}$. In the other models, the effect of the $\sigma$ meson pole term gives positive contribution to the scattering length, namely it is attractive at the threshold. In all cases, the scattering length is slightly enhanced through the unitarization procedure. The slightly small result in model B may indicate that the pure $\pi\pi$ molecule for the description of the physical $\sigma$ meson is not sufficient. In the models A, C and D, the scattering length is enhanced by the sigma pole term. In the chiral perturbation theory, the next-to-leading order terms, dominated by the $\rho$ exchange contribution, are responsible to reproduce the experimental value of the scattering length. In spite of the simple model setup, we find that the results are in reasonable agreement with data; the values of the scattering length are not very far from the experimental determination and the low energy theorem.

Next we study the pole structure of the unitarized amplitude in each model in vacuum. The result of the pole positions and their possible origins are summarized in Tables~\ref{tbl:model} and \ref{tbl:modeldata}. For reference, we note that the pole position is $z=441-272i$ MeV in the recent analysis of Ref.~\cite{Caprini:2005zr}. In models A and C, the bare $\sigma$ meson acquires finite width through the coupling to $\pi\pi$ state, and one pole appears in the amplitude. Thus the origin of the pole stems from the bare $\sigma$ pole in the tree-level amplitude. In model B, the attractive $\pi\pi$ interaction generates a resonance in the amplitude, which can be interpreted as the $\sigma$ meson. Note that this is a resonance in $s$-wave, which cannot appear in a simple nonrelativistic potential model. The energy dependence of the interaction is essential for the dynamical generation of an $s$-wave resonance. The energy dependence of the chiral interaction also plays an important role for the double-pole structure of the $\Lambda(1405)$ baryon resonance~\cite{Jido:2003cb,Hyodo:2007jq}, since the lower energy pole of the $\Lambda(1405)$ originates in the $s$-wave resonance of the $\pi \Sigma$ channel. 

Although our purpose here is to prepare the models with purified structure of the $\sigma$ meson, it is possible to phenomenologically tune the subtraction constant from the value in Eq.~\eqref{eq:ampi} for more quantitative description in model B. For instance, choosing the phenomenological subtraction constant $a^{(1)}_{\text{pheno}}=-5.75$, we obtain the empirical scattering length of Ref.~\cite{Batley:2007zz,BlochDevaux:2009zzb} as $a=0.222 m_{\pi}^{-1}$. In this case, the pole position is $z=469-179i$ MeV. When we use the subtraction constant $a^{(2)}_{\text{pheno}}=-4.5$, scattering length is $a=0.204 m_{\pi}^{-1}$ and the pole position of the amplitude is at $z=463-249i$ MeV, which is close to the recent determination~\cite{Caprini:2005zr}. However, it is shown in Ref.~\cite{Hyodo:2008xr} that the deviation of the subtraction constant from the value in Eq.~\eqref{eq:ampi} is equivalent to the introduction of the pole term in the interaction kernel. Indeed, following Ref.~\cite{Hyodo:2008xr}, we obtain the effective mass of the pole term as
\begin{equation}
    M_{\text{eff}}
    =\sqrt{-\frac{16\pi^2\cond{\sigma}^2}{\Delta a}+\frac{m_{\pi}^2}{2}},
    \quad \Delta a = a_{\text{pheno}}+\frac{\pi}{\sqrt{3}},
    \label{eq:Meff}
\end{equation}
which gives us $M_{\text{eff}}^{(1)}=597$ MeV for $a^{(1)}_{\text{pheno}}$ and $M_{\text{eff}}^{(2)}=720$ MeV for $a^{(2)}_{\text{pheno}}$. Therefore, to examine the purely dynamically generated $\sigma$ meson, we use Eq.~\eqref{eq:ampi} in model B. We consider model C as a representative for the cases with pole term in the interaction kernel.

In model D ($x=1/2$), we obtain two resonance poles in the amplitude. We are tempted to interpret that one of the poles originates in the bare $\sigma$ state, and another is generated dynamically by $\pi\pi$ interaction, as discussed in Ref.~\cite{Yokokawa:2002pw}. This interpretation may be reasonable from the sign of the contact interaction. In models A and C, the contact interaction is repulsive,\footnote{As pointed out in Ref.~\cite{Ishida:1995xx,Ishida:1996fp}, the four-pion interaction in the linear sigma model is repulsive. In section 2.2, we show that this also holds including the contributions from the $s$-wave projection of $u$- and $t$-channel diagrams.} and hence, no state is dynamically generated. In model D, the bare state induces one pole, and the attractive force in the contact interaction generates an additional pole.

\begin{figure}[tbp]
    \centering
    \includegraphics[width=6.5cm,clip]{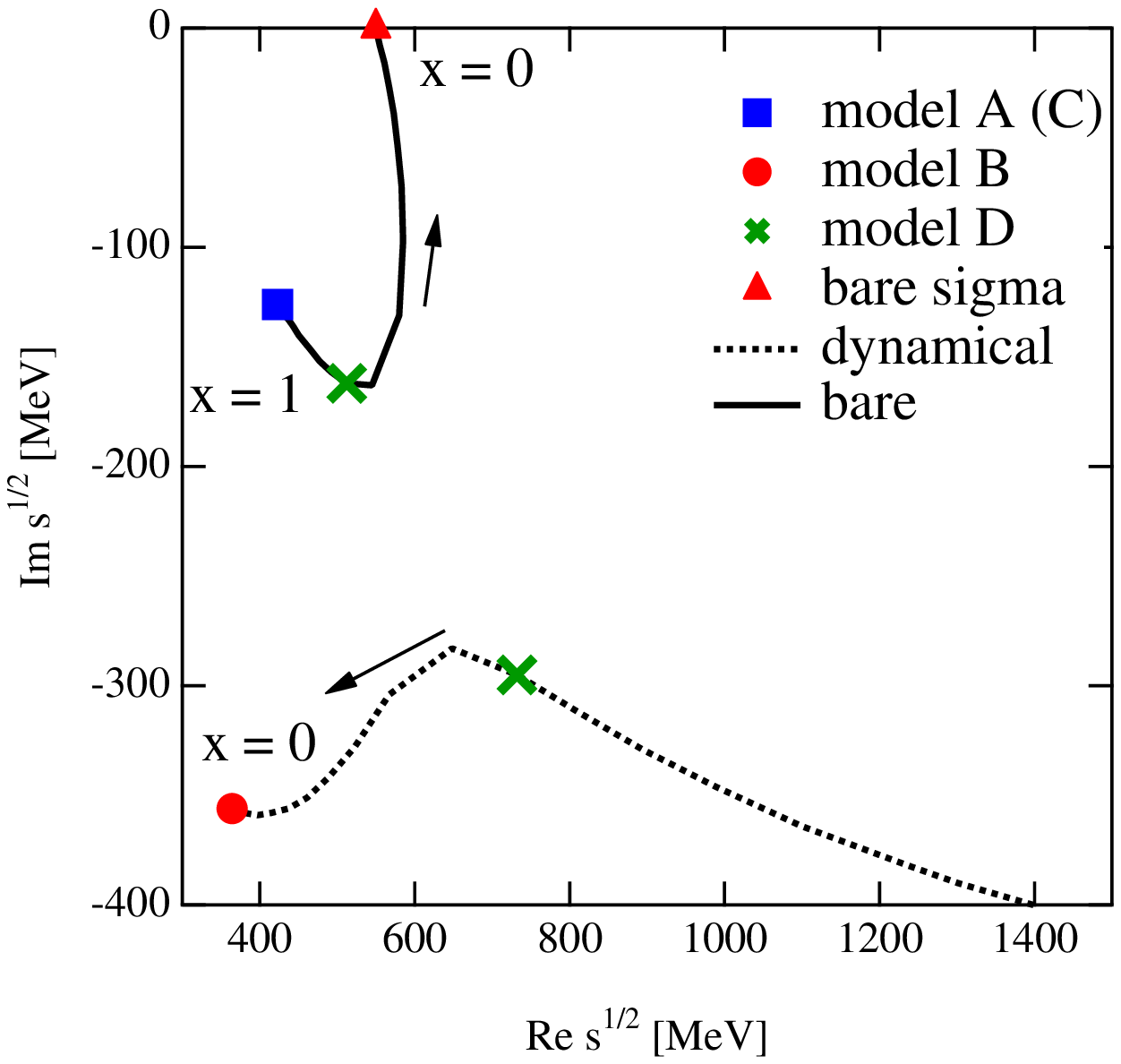}
    \includegraphics[width=6.5cm,clip]{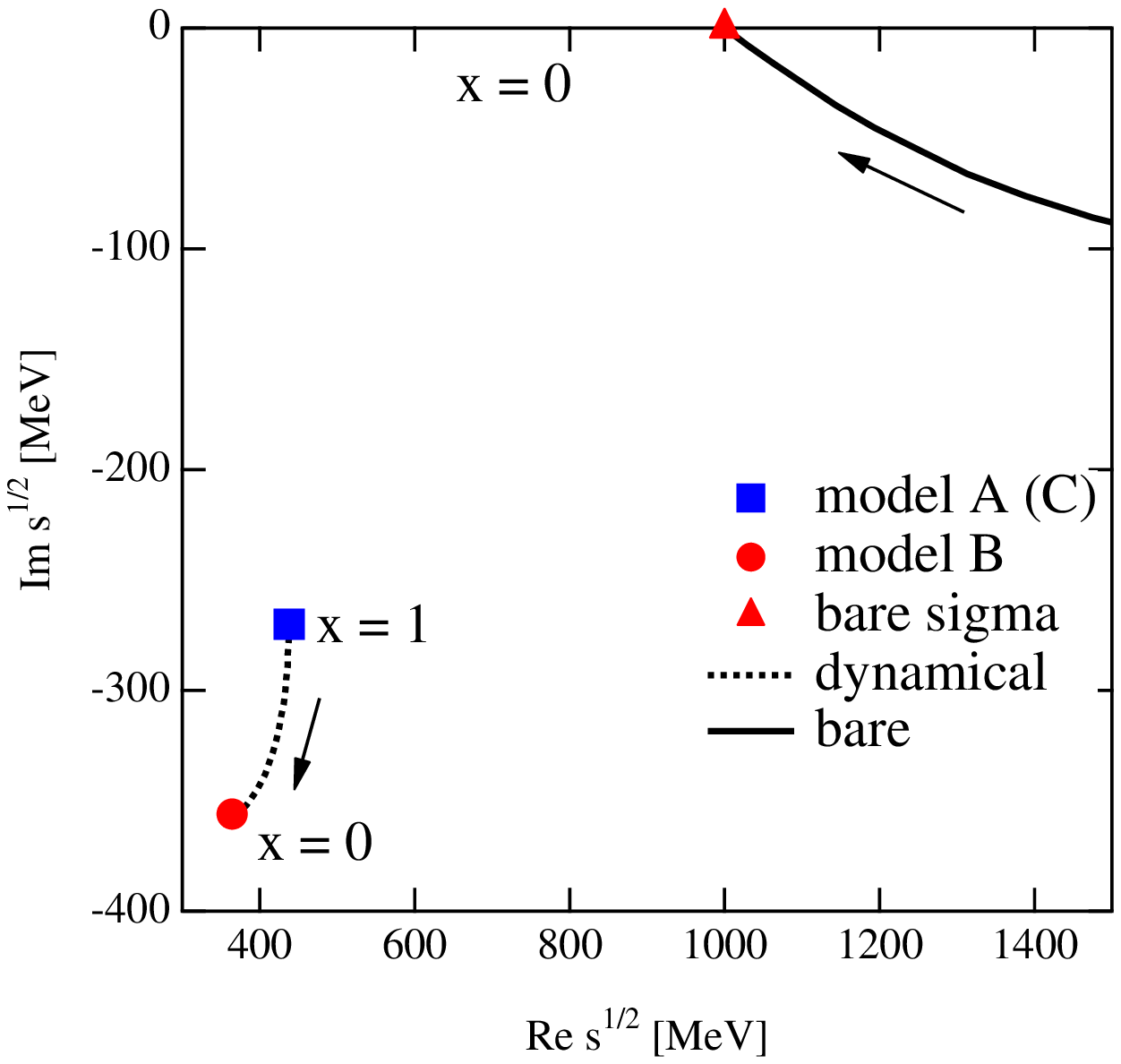}
    \caption{\label{fig:modelext}
    Left: Trajectory of the pole positions of the scattering amplitude 
    varying the parameter $x$. Arrows indicate the direction of the 
    movement of the pole as the parameter $x$ is decreased from 1 to 0.
    Right: the same plot with bare $\sigma$ mass $m_{\sigma}=1$ GeV.
    }
\end{figure}

To further illustrate the origin of the poles in model D, we study the trajectory of the pole positions by varying the parameter $x$ in the left panel of Fig.~\ref{fig:modelext}. The pole obtained in models A and C is plotted by the square, which moves toward the bare $\sigma$ pole (denoted by the triangle) as we decrease the parameter $x$ from 1 to 0. Although the bare pole is decoupled from the interaction in the limit of $x=0$, the pole position asymptotically approaches the mass of the bare $\sigma$ pole. The decrease of the parameter $x$ corresponds to the suppression of the coupling of the bare pole to the scattering state. This fact therefore implies that the origin of the pole in models A and C is attributed to the bare $\sigma$ pole. On the other hand, the pole obtained in model B (circle) moves to the higher energy region and finally disappears when we increase $x$ from 0 to 1. It is natural to interpret this pole as a dynamical state generated in the $\pi\pi$ attraction. Since the attractive component of the contact interaction gradually switched off as $x\to 1$, the resonance becomes loosely bound and finally dissolves into continuum.

The two poles in model D (crosses in Fig.~\ref{fig:modelext}) are on the two trajectories, one connecting the pole at $x=1$ with the bare pole, and the other from the pole at $x=0$ in model B. Based on the above discussion, we consider the former trajectory as the ``bare state'' branch, and the latter trajectory as the ``dynamically generated state'' branch. Thus, we may identify one of the poles in model D originates from the bare $\sigma$ state, while the other is dynamically generated by the $\pi\pi$ attraction. 

It is instructive to examine the case with a large value for the bare $\sigma$ mass in vacuum. In the right panel of Fig.~\ref{fig:modelext}, we show the pole trajectories for $x:0\to 1$ with the bare $\sigma$ mass as $m_{\sigma}=1$ GeV where qualitatively different pattern from the left panel emerges. The pole at $x=1$ moves to the dynamically generated pole at $x=0$, not to the bare $\sigma$ pole. The other branch from the bare pole is connected to infinity. In this case, the energy of the bare $\sigma$ pole (1 GeV) is very high compared with the pole in the amplitude. The propagator of the bare $\sigma$ pole is then regarded as an effective contact interaction, which supplies the attractive force for the relevant energy region:
\begin{align}
    T_{\text{tree}}^{\text{(pole)}}(s;x)
    =&3x 
    \frac{(m_{\sigma}^{2}-m_{\pi}^2)^2}{\cond{\sigma}^2}\frac{1}{m_{\sigma}^2}
    \left(1+\frac{s}{m_{\sigma}^2}+\dotsb\right) 
    \label{eq:polecontract} \quad \text{for}\quad s\ll m_{\sigma}^2.
\end{align}
If the bare mass of the $\sigma$ is sufficiently high, the tree level amplitude is effectively given by the leading order contribution of Eq.~\eqref{eq:TtreeNL} plus some higher order corrections. In this way, the origin of the pole at $x=1$ is considered as the dynamically generated one, when $m_{\sigma}$ is taken to be 1 GeV. We should however keep in mind that the property of a pole may change as we vary the parameter $x$.\footnote{The character change of the state during the extrapolation was discussed in the study of the $N_c$ dependence of the $\Lambda$ resonances in Refs.~\cite{Hyodo:2007np,Roca:2008kr}.} In addition, the dynamically generated state and the bare states are not orthogonal, so the physical state is the mixture of both. In this sense, the analysis of the pole trajectories should be regarded as a guidance for the origin of the state. 

To summarize, we prepare four models in which the amplitudes have different origins of poles as shown in Table~\ref{tbl:model}. We consider the case of the dynamically generated $\sigma$ meson (model B), the case of the $\sigma$ meson as the CDD pole (model A for the chiral partner of the pion and model C for the state generated by quark-gluon dynamics), and the mixture of both dynamical and CDD pole  (model D). 

\subsection{Softening of the $\sigma$ meson}\label{subsec:softening}

For each model, we calculate the scattering amplitude, varying the chiral condensate $\cond{\sigma}$ from $\cond{\sigma}_0$ to 0. The spectrum of the $\pi\pi$ scattering amplitude is presented by the reduced cross section
\begin{equation}
    \bar{\sigma}=\frac{|T|^2}{s} .
    \label{eq:cross}
\end{equation}
We also plot the trajectory of the pole position of the amplitude to visualize the effect of the symmetry restoration. For convenience, plots are given by the functions of the total center-of-mass energy $\sqrt{s}$.

\begin{figure}[tbp]
    \centering
    \includegraphics[width=6.5cm,clip]{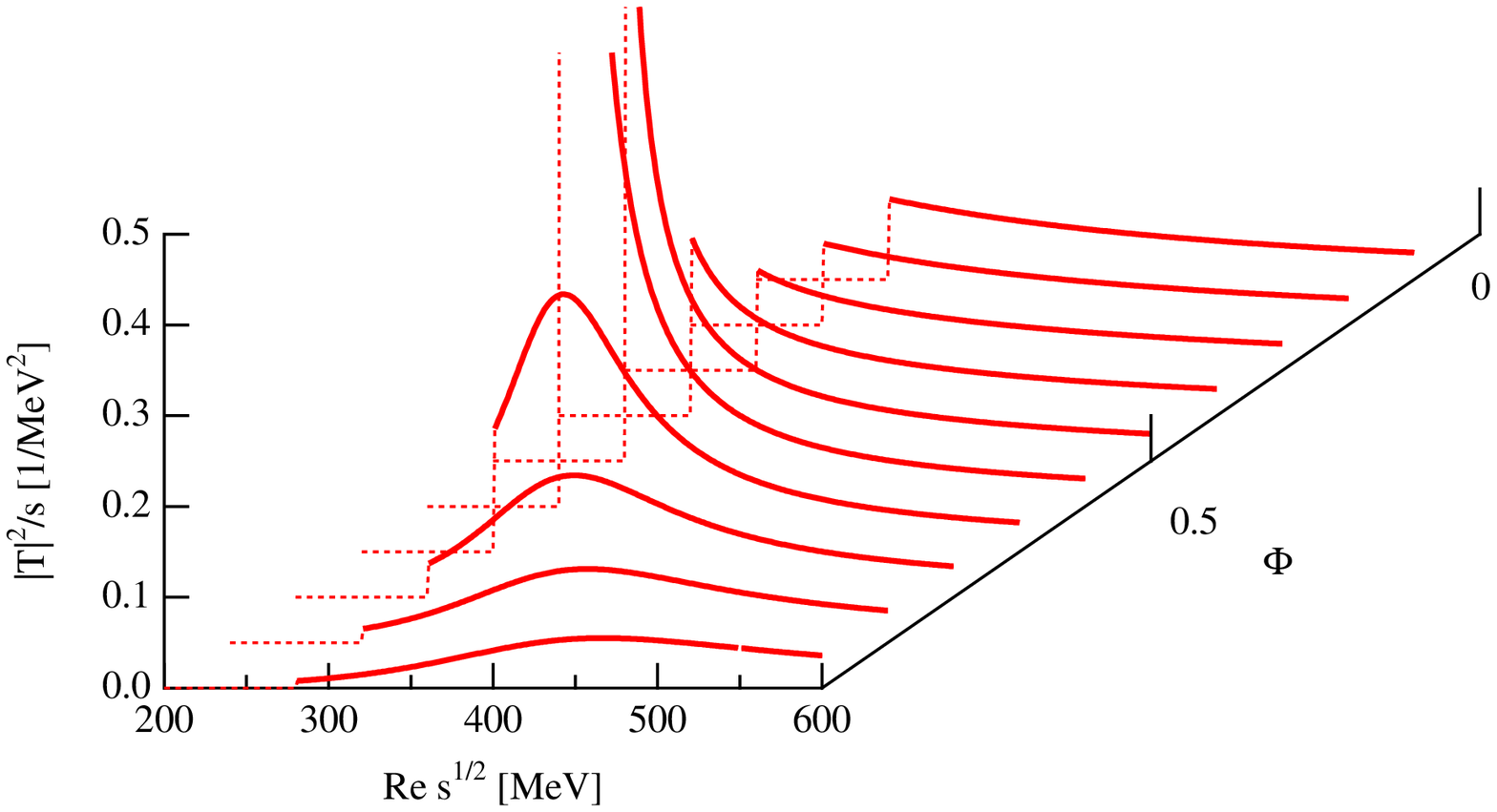}
    \includegraphics[width=6.5cm,clip]{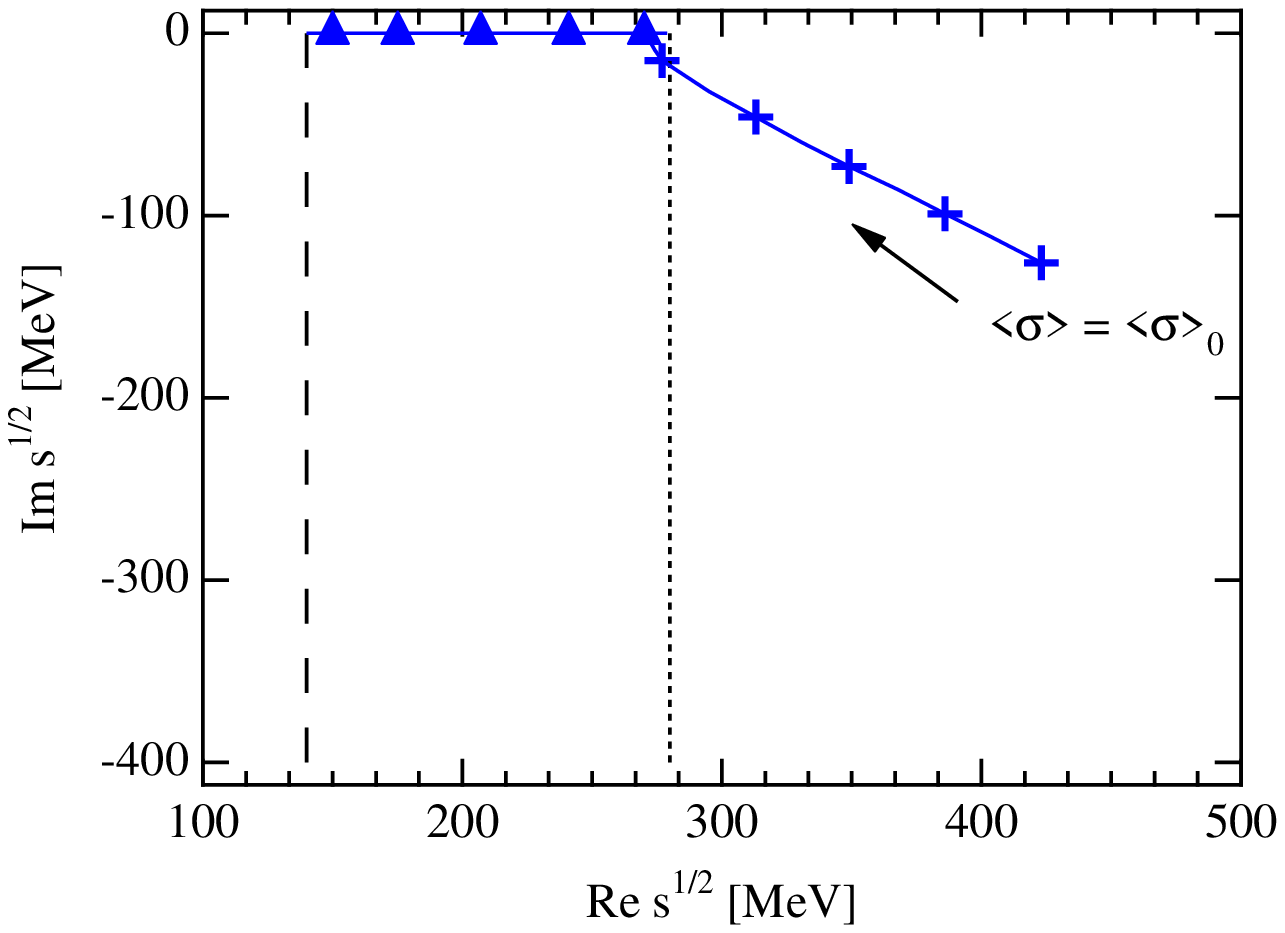}
    \caption{\label{fig:modelA}
    Spectra of the $\sigma$ meson (left)
    and the trajectory of the pole positions (right) in model A 
    ($x=1$, case I). 
    The symbols are marked with each 0.1 step of 
    $\Phi=\cond{\sigma}/\cond{\sigma}_0$.
    The arrow indicates the direction of the movement of the pole as 
    the condensate $\cond{\sigma}$ is decreased from $\cond{\sigma}_0$ 
    to 0.
    The poles on the first Riemann sheet is denoted by triangles, 
    while the poles on the second Riemann sheet is plotted by crosses.
    The dotted (dashed) line represents the energy of the threshold 
    (mass of the pion).}
\end{figure}

Fig.~\ref{fig:modelA} shows the results of model A which corresponds to the standard linear sigma model. We observe that the softening of $\sigma$ takes place. The pole of the $\sigma$ moves to the lower energy side with reducing its width, and finally it becomes a bound state below the threshold at $\cond{\sigma}\sim 0.6\cond{\sigma}_0$. Around this value of $\cond{\sigma}$, the bare mass of the $\sigma$ also moves to the bound region below the threshold. The spectrum of the $\sigma$ meson shows a clear peak structure around the threshold. In the limit of $\cond{\sigma}\to 0$, the pole approaches the mass of the pion and finally coincide with it. Since the sigma pole moves far away from the threshold, with reducing the coupling strengths to the $\pi\pi$ state, the spectrum observed above the threshold shows no prominent structure for $\cond{\sigma}\leq 0.3\cond{\sigma}_0$. This is consistent with the behavior studied in section~\ref{subsec:limit}, although we should keep in mind that the effect of the left hand cut, which we have neglected, would become important for the bound state below threshold. This model manifests the standard scenario of the softening where the movement of the pole is driven by the decrease of the bare mass of the $\sigma$ pole in the interaction kernel.\footnote{Around the threshold, there is a small region in which the virtual state is formed as in model B. This reflects the effect of the change of the property of the pole, as we discuss for model C.}

\begin{figure}[tbp]
    \centering
    \includegraphics[width=6.5cm,clip]{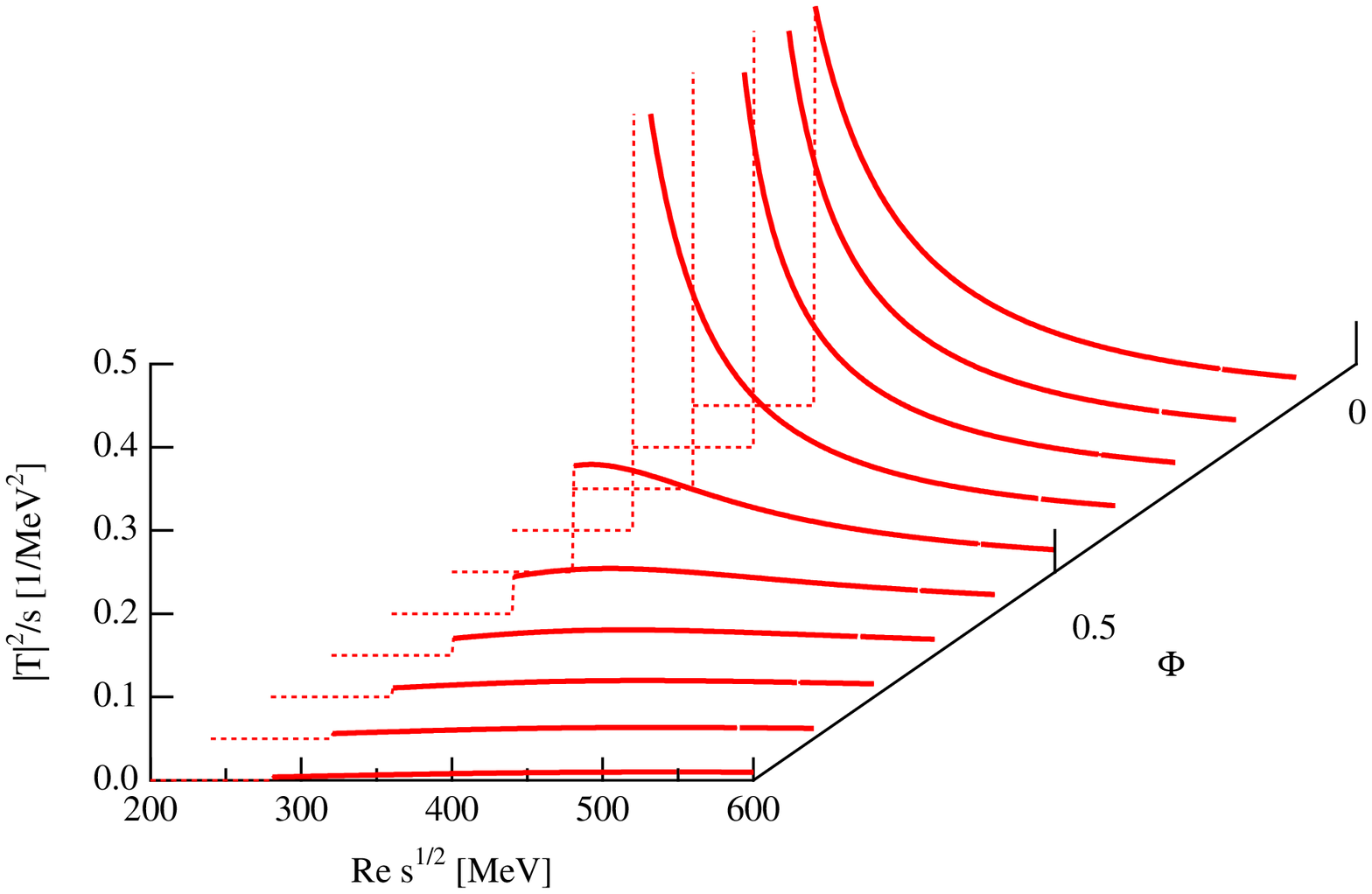}
    \includegraphics[width=6.5cm,clip]{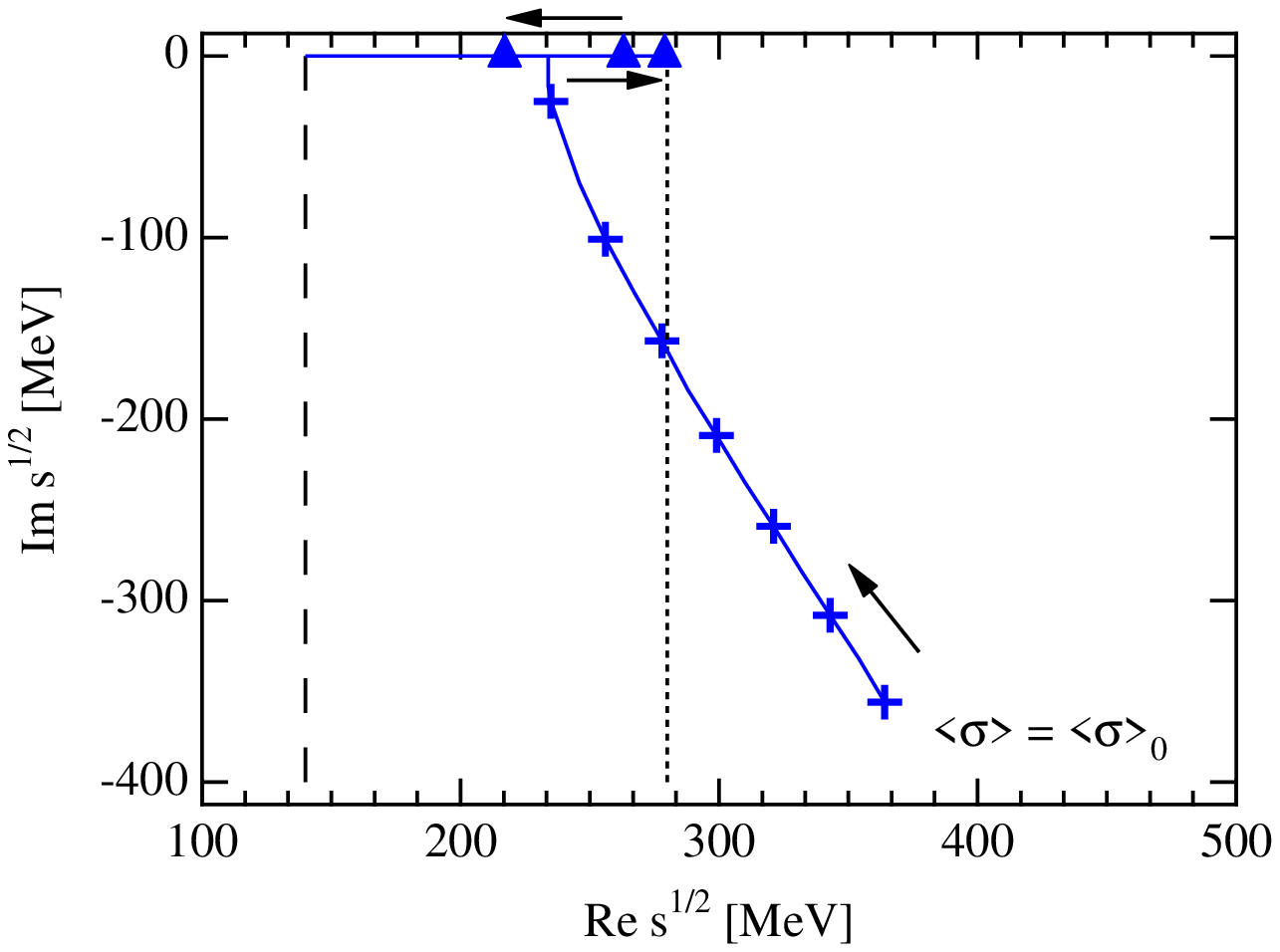}
    \caption{\label{fig:modelB}
    Spectra of the $\sigma$ meson (left)
    and the trajectory of the pole positions (right) in model B ($x=0$). 
    The symbols are marked with each 0.1 step of 
    $\Phi=\cond{\sigma}/\cond{\sigma}_0$.
    The arrows indicate the direction of the movement of the pole as 
    the condensate $\cond{\sigma}$ is decreased from $\cond{\sigma}_0$ 
    to 0.
    The poles on the first Riemann sheet is denoted by triangles, 
    while the poles on the second Riemann sheet is plotted by crosses.
    The dotted (dashed) line represents the energy of the threshold 
    (mass of the pion).}
\end{figure}

The results of model B is shown in Fig.~\ref{fig:modelB}, where the pole is dynamically generated by the attractive $\pi\pi$ interaction. In this case, the change of the spectrum as well as the trajectory of the pole are qualitatively different from those of model A. We observe that the pole moves below the threshold keeping the finite width~\cite{FernandezFraile:2007fv}. This phenomena is caused by the appearance of the virtual state. It is known that when the attractive interaction is strengthened, an $s$-wave resonance can become a virtual state which is characterized by the pole on the second Riemann sheet below the threshold energy. In model B, the reduction of the chiral condensate results in the enhancement of the attractive interaction as seen in Eq.~\eqref{eq:TtreeNL}, and hence the resonance in vacuum turns into a virtual state, before the two-body $\pi\pi$ system forms the bound state.

Because of this special nature of an $s$-wave resonance, the change of the spectrum shows a different pattern from the softening of model A in Fig.~\ref{fig:modelA}. In model A, the peak of the $\sigma$ meson becomes sharp and it causes the divergence of the spectral function when the real part of the pole approaches the threshold. On the other hand, the pole of the dynamically generated $\sigma$ meson first moves to the second Riemann sheet of lower energy region than the threshold. In this case, due to the finite width, the spectrum does not shows the prominent peak structure, when the real part of the pole crosses the threshold. Once the pole reaches the real axis it moves toward the threshold on the second Riemann sheet,\footnote{There are always two poles in the amplitude, namely, there is another branch of the pole trajectory in addition to the trajectory shown in Fig.~\ref{fig:modelB}. When the pole has finite imaginary part, the other pole exists at $z=z^*$ with $\im z>0$ on the second Riemann sheet. After the pole reaches the real axis, the other pole goes to the lower energy direction on the second Riemann sheet. Here we focus on the most relevant pole to the spectrum above the threshold.} and finally it becomes a bound state on the first Riemann sheet, where we observe the divergence of the spectral function at the threshold. Since the interaction kernel given in Eq.~\eqref{eq:TtreeNL} is a monotonically increasing function of $s$, we can use the argument in Refs.~\cite{Hyodo:2006yk,Hyodo:2006kg} to define the critical coupling strength with which the two-body attractive interaction generates a bound state. In the present case, the decay constant is changed with the coupling strength being fixed. The critical value of the condensate is
\begin{align}
    \cond{\sigma}
    =\sqrt{\frac{7}{2\sqrt{3}\pi}}
    \frac{m_{\pi}}{4}
    \sim 0.3\cond{\sigma}_0
    \label{eq:critical} .
\end{align}
This is indeed the value of $\cond{\sigma}$ where the pole becomes the bound state. In the end, the $\sigma$ pole is degenerated with the pion mass for $\cond{\sigma}\to 0$ in agreement with the discussion in section~\ref{subsec:limit}. The fate of the dynamically generated $\sigma$ meson has been discussed without explicit symmetry breaking in Ref.~\cite{Yokokawa:2002pw}. 

We note that it is essential to introduce the finite pion mass for the appearance of the virtual state, since the virtual state can appear in the energy region below the threshold $\sqrt{s}<2m_{\pi}$. The mechanism of the appearance of the virtual state is the same as the quark mass dependence of the $\sigma$ pole found in Ref.~\cite{Hanhart:2008mx}. In this respect, inclusion of the finite pion mass should not change the softening of the $\rho$ meson in $I=J=1$ channel from the results in the chiral limit~\cite{Yokokawa:2002pw}, since the $\rho$ appears in $p$-wave $\pi\pi$ amplitude and thus no virtual state is allowed (see also Fig. 1 in Ref.~\cite{Hanhart:2008mx}). In this sense, for $m_{\pi}\neq 0$, the softening of the dynamically generated $\sigma$ is qualitatively different from that of the $\rho$, in contrast to the universality found in the chiral limit~\cite{Yokokawa:2002pw}. This is indeed demonstrated in Ref.~\cite{FernandezFraile:2007fv}.

It is instructive to study the pole structure in the different Riemann sheet for the elementarity/compositeness of the sigma meson. The old discussion of the compositeness by Weinberg~\cite{Weinberg:1962hj,Weinberg:1963zz,Weinberg:1965zz} was later interpreted as the asymmetry of the poles in the first and the second Riemann sheets~\cite{Morgan:1992ge,Baru:2003qq}; if a bound state is an elementary (a composite) particle, the shadow pole in the second Riemann sheet locates close to (far away from) the position of the bound state pole in the first Riemann sheet. To study the structure of the bound states in models A and B further, we search for the shadow pole in the second Riemann sheet when the bound state appears just below the threshold. The results are summarized in Table~\ref{tbl:pole}, together with the value of $\Phi=\cond{\sigma}/\cond{\sigma}_0$ to have the bound state at $\sqrt{s}=279$ MeV. The virtual pole appears at $240$ MeV in model A, while the virtual pole in model B is at $140$ MeV which is far away from the bound state pole at the threshold. This result indicates the bare $\sigma$ nature of the bound state in model A, and the bound state in model B can be interpreted as the $\pi\pi$ molecule dominant state.
\begin{table}[tbp]
    \centering
    \caption{Pole structure of the $\sigma$ meson in models A, B and C, when the bound state appears just below the threshold. Bound state pole is on the first Riemann sheet, while the virtual state pole lies in the second Riemann sheet.}
    \begin{tabular}{c|ccc}
    \hline
     & $\Phi=\cond{\sigma}/\cond{\sigma}_0$ & bound state [MeV] & virtual state [MeV]  \\
    \hline
    model A & $0.55$ & $279$ & $240$   \\
    model B & $0.28$ & $279$ & $140$   \\
    model C & $0.32$ & $279$ & $150$   \\
    \hline
    \end{tabular}
    \label{tbl:pole}
\end{table}

\begin{figure}[tbp]
    \centering
    \includegraphics[width=12cm,clip]{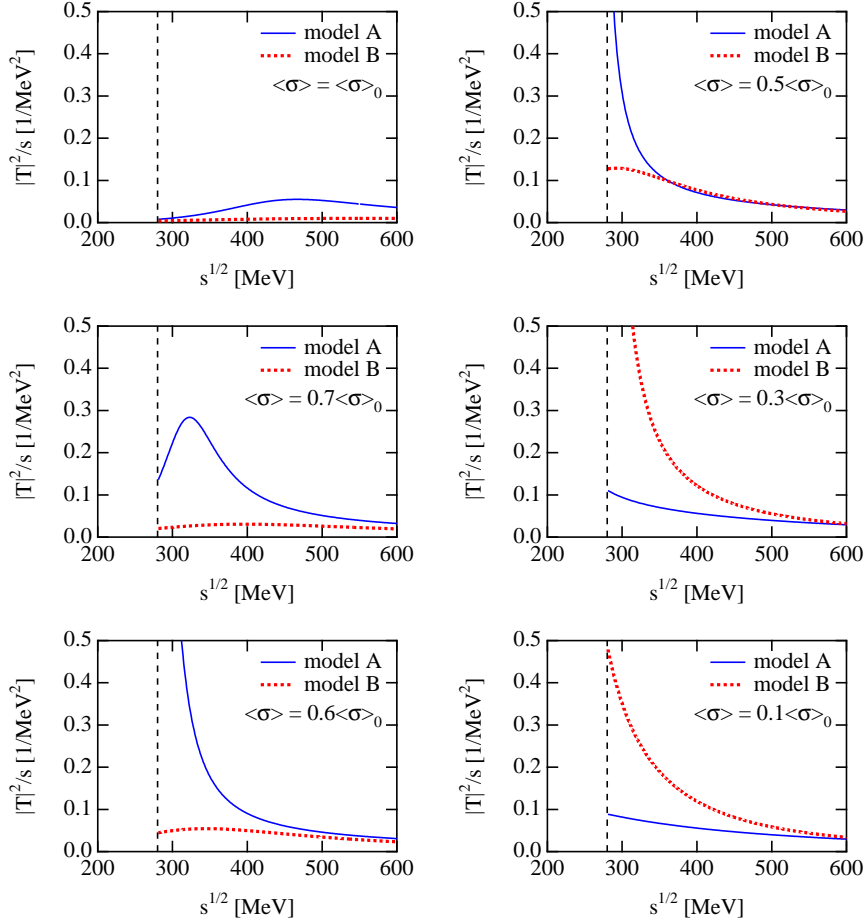}
    \caption{\label{fig:modelAB}
    Comparison of spectra of the $\sigma$ meson 
    in model A (solid lines) and model B (dotted lines) 
    for several values of the condensate $\cond{\sigma}$.
    The dashed line represents the threshold.}
\end{figure}

In Fig.~\ref{fig:modelAB} we compare the spectra of model A and model B where the difference between two models is clear. In model A, the peak of the $\sigma$ spectrum becomes sharp at $\cond{\sigma}=0.6\cond{\sigma}_0$, while no prominent structure can be seen in model B, although the real part of the pole is close to the threshold, as seen in right panel of Fig.~\ref{fig:modelB}. Because of the finite width, the pole in model B does not affect the spectrum on the real axis very much. The strong threshold enhancement in model B is observed at $\cond{\sigma}=0.3\cond{\sigma}_0$, where the peak of the model A is already flattened. Thus, the threshold enhancement takes place at different values of $\cond{\sigma}$ in model A and model B, because the effect is caused by different mechanism. In model A, the softening is driven by the movement of the bare $\sigma$ pole. In model B it is caused by the enhancement of the $\pi\pi$ attractive interaction and the formation of the virtual state is crucial for the difference from model A.

Near the chiral restoration $\cond{\sigma}=0.1\cond{\sigma}_0$, model B still shows the strong peak at the threshold, while the strength in model A is rather weak. This is partly caused by the difference of the movement of the pole as seen in Figs.~\ref{fig:modelA} and \ref{fig:modelB}, but is also related to the asymptotic behavior of the $\sigma\pi\pi$ coupling. In the restoration limit, the coupling vanishes in model A as seen in Eq.~\eqref{eq:limitcaseI}, while it remains finite in model B as in Eq.~\eqref{eq:NLScoupling}. Therefore, although the $\sigma$ pole moves toward the pion mass in both models, the pole in model B has stronger effect on the $\pi\pi$ spectrum above the threshold. In this way, the dynamically generated $\sigma$ meson in model B shows the threshold enhancement of the spectrum, but its behavior is qualitatively different from the $\sigma$ meson as the chiral partner.

\begin{figure}[tbp]
    \centering
    \includegraphics[width=6.5cm,clip]{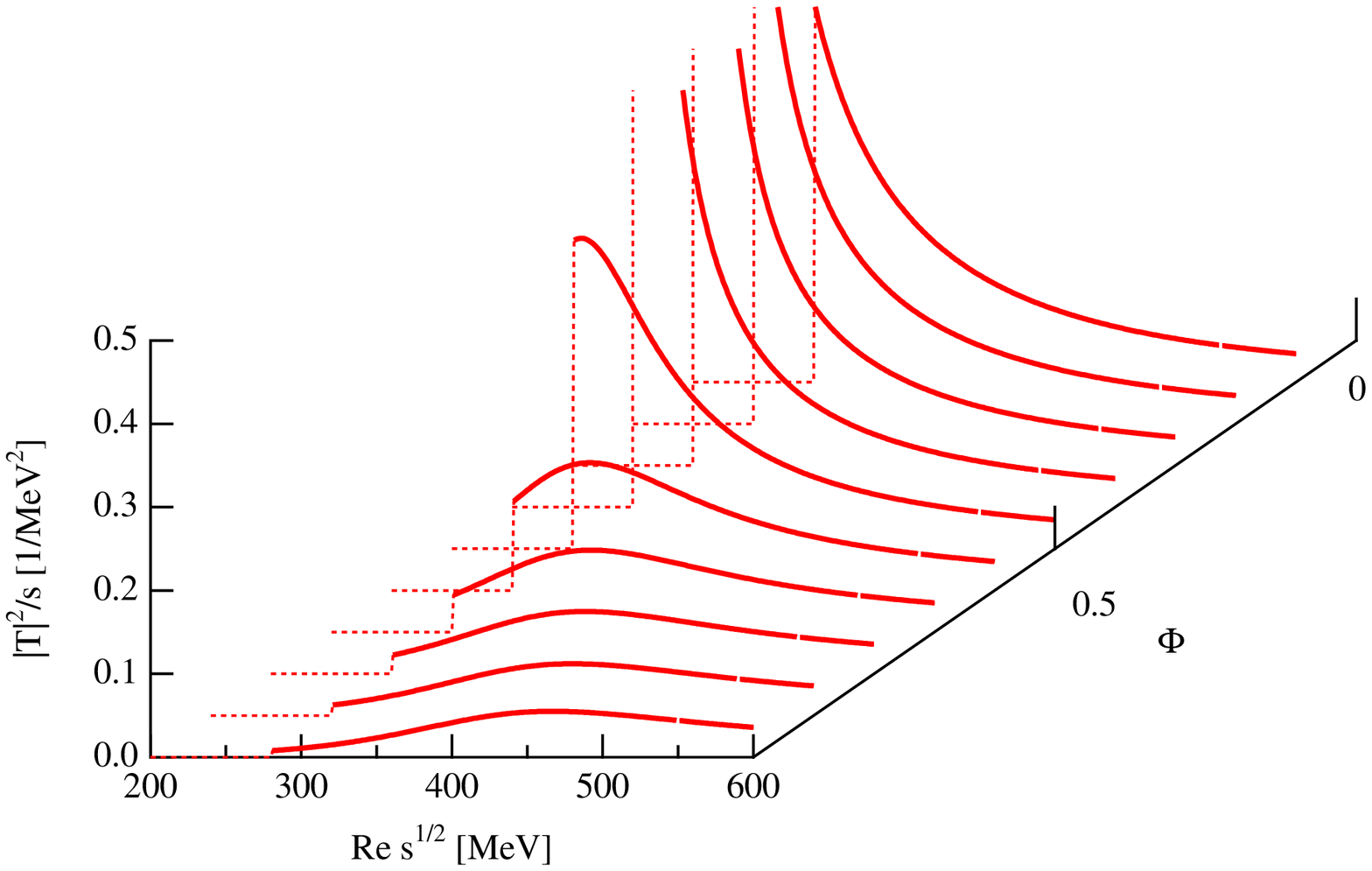}
    \includegraphics[width=6.5cm,clip]{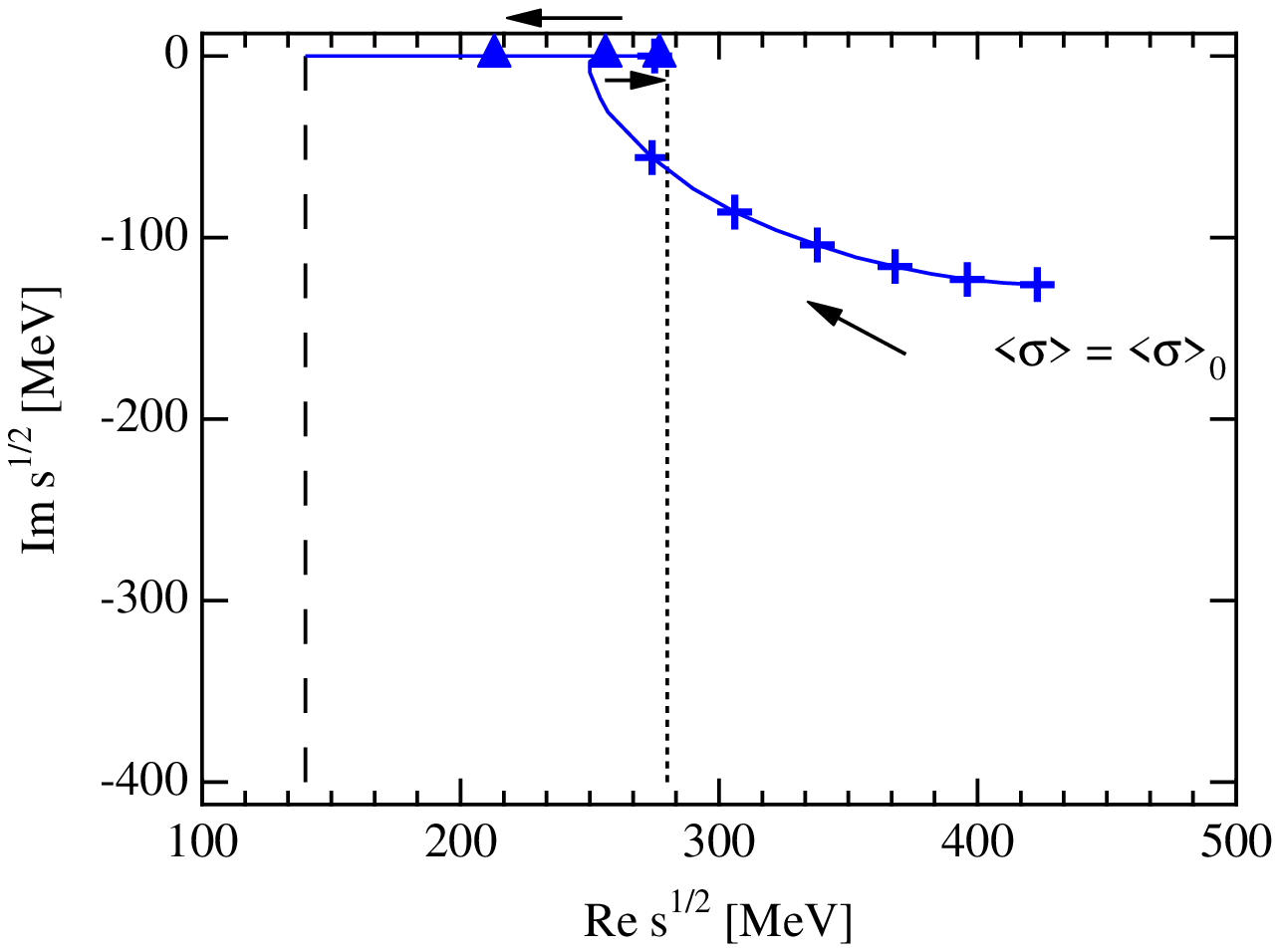}
    \caption{\label{fig:modelC}
    Spectra of the $\sigma$ meson (left)
    and the trajectory of the pole positions (right) in model C 
    ($x=1$, case II).
    The symbols are marked with each 0.1 step of 
    $\Phi=\cond{\sigma}/\cond{\sigma}_0$.
    The arrows indicate the direction of the movement of the pole as 
    the condensate $\cond{\sigma}$ is decreased from $\cond{\sigma}_0$ 
    to 0.
    The poles on the first Riemann sheet is denoted by triangles, 
    while the poles on the second Riemann sheet is plotted by crosses.
    The dotted (dashed) line represents the energy of the threshold 
    (mass of the pion).}
\end{figure}

In Fig.~\ref{fig:modelC}, we show the result of model C. In this model, the interaction contains the bare $\sigma$ pole whose mass does not change with the symmetry restoration. Qualitative behavior of the pole position is similar to that in Fig.~\ref{fig:modelB}, namely, the pole becomes the virtual state before forming the bound state. In comparison with models A and B, this indicates that the resonance is dynamically generated. However, as we have discussed in section~\ref{subsec:vacuum}, the origin of the pole in vacuum is attributed to the bare pole in this model. This implies that the nature of the resonance is changing from the CDD pole to the dynamically generated one, as the symmetry is gradually restored.

\begin{figure}[tbp]
    \centering
    \includegraphics[width=6.5cm,clip]{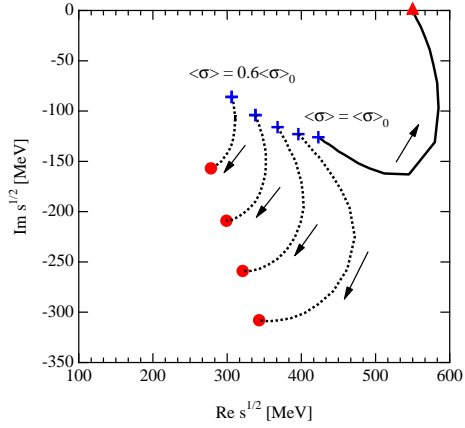}
    \caption{\label{fig:modelC_detail}
    Pole positions of the $\sigma$ meson in model C 
    for $\cond{\sigma}=\cond{\sigma}_0$ to $0.6\cond{\sigma}_0$ (crosses). 
    The trajectory from each pole is drawn
    by changing the parameter $x$ from 1 to 0.
    At $x=0$, the dynamically generated poles are represented by circles,
    while the bare pole is denoted by the triangle.}
\end{figure}

Actually, the change of the property of the $\sigma$ pole can be traced by studying the behavior of the pole in the limit of $x\to 0$. Fig.~\ref{fig:modelC_detail} shows the trajectory of the pole when the parameter $x$ is changed from 1 to 0, for several values of $\cond{\sigma}$. As we saw in section~\ref{subsec:vacuum}, for $\cond{\sigma}=\cond{\sigma}_0$, by the decrease of the parameter $x$, the pole approaches the energy of the bare state. For the smaller values of $\cond{\sigma}$, the pole moves toward the position of dynamically generated pole at $x=0$. This indicates that the property of the pole changes from the bare pole origin to the dynamically generated one. This change can be understood in the following way. When we decrease the condensate $\cond{\sigma}$, the pole in the amplitude moves to the lower energy region, so the relative importance of the bare pole contribution decreases and it is effectively regarded as an attractive contact interaction given in Eq.~\eqref{eq:polecontract}. As a consequence, the property of the $\sigma$ pole in Fig.~\ref{fig:modelC} is dominated by the dynamically generated component when its real part crosses the threshold, leading to the formation of the virtual state. The dynamical nature of the $\sigma$ pole near threshold can be further confirmed by checking the position of the shadow pole (see Table~\ref{tbl:pole}). We find the virtual pole at 150 MeV for the bound state at 279 MeV. The large deviation indicates the composite nature of the bound state.

The result in model C is also instructive in comparison with model A and model B. We have discussed the difference of the softening between model A and model B, but it should be noted that the pole positions in vacuum are different from each other. In this respect, model C is a good example which has the same amplitude and the pole position with model A in vacuum, and shows the softening pattern of dynamically generated sigma meson for small $\cond{\sigma}$. Indeed, through the argument in Ref.~\cite{Hyodo:2008xr}, model C can be also regarded as the model in which the interaction has no bare pole term with the subtraction constant being adjusted such that the pole position in vacuum becomes the same as those in model A. Comparing model C with model A, we conclude that the position of the pole in vacuum does not change the qualitative feature of the softening of the dynamically generated sigma.

\begin{figure}[tbp]
    \centering
    \includegraphics[width=6.5cm,clip]{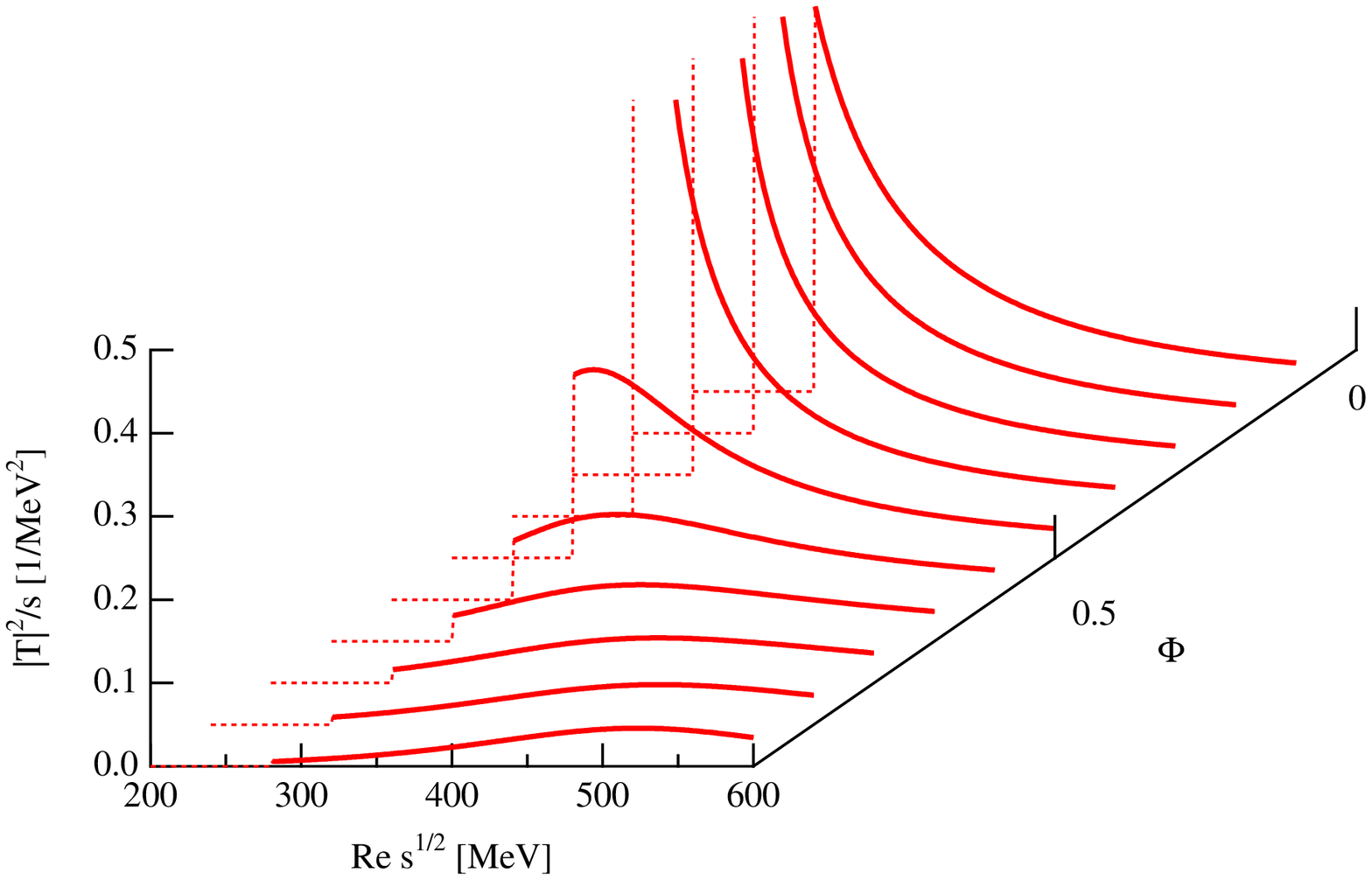}
    \includegraphics[width=6.5cm,clip]{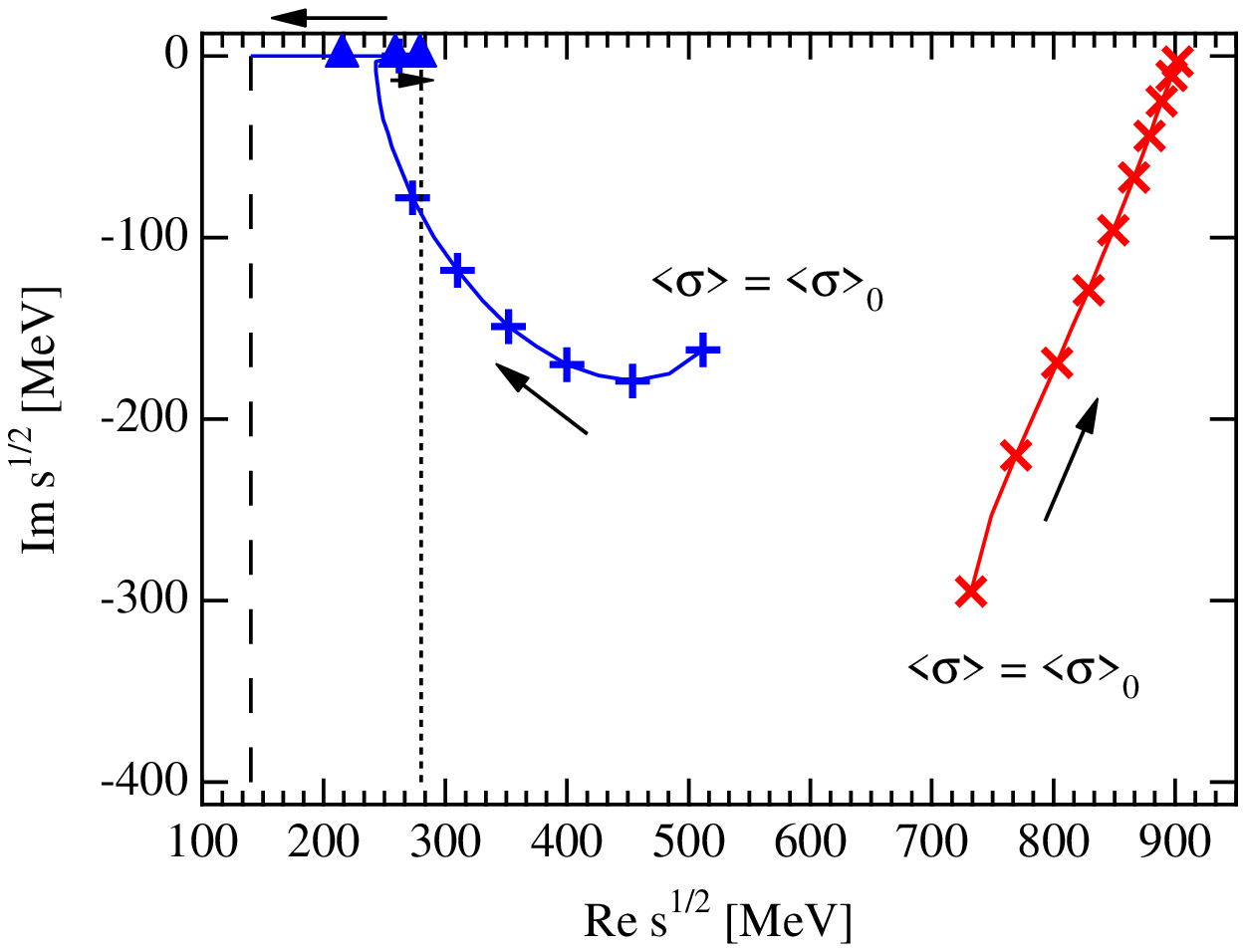}
    \caption{\label{fig:modelD}
    Spectra of the $\sigma$ meson (left)
    and the trajectory of the pole positions (right) in model D 
    ($x=1/2$, case II).
    The symbols are marked with each 0.1 step of 
    $\Phi=\cond{\sigma}/\cond{\sigma}_0$.
    The arrows indicate the direction of the movement of the pole as 
    the condensate $\cond{\sigma}$ is decreased from $\cond{\sigma}_0$ 
    to 0.
    The poles on the first Riemann sheet is denoted by triangles, 
    while the poles on the second Riemann sheet is plotted by crosses.
    The dotted (dashed) line represents the energy of the threshold 
    (mass of the pion).}
\end{figure}

Finally we show the result of model D in Fig.~\ref{fig:modelD} where two poles appear in vacuum. As the symmetry is restored, the lower energy pole goes toward the threshold and shows the similar pattern with models B and C. Although the origin of the lower energy pole in model D is considered to be the bare pole in vacuum, the similar character change with model C takes place and the nature of the pole becomes dynamically generated one when the pole comes close to the $\pi\pi$ threshold.

The higher energy pole in vacuum moves to the higher energy direction with 
reducing its width. At $\cond{\sigma}\to 0$, the pole approaches $\sqrt{s}\sim 903$ MeV. This is not the position of the bare pole, but is the zero of the interaction $T_{\text{tree}}$. Note that the interaction kernel has a zero only for $C<x<1$, where the contact interaction changes the sign (see Table~\ref{tbl:contact}). For an illustration, let us consider that $T_{\text{tree}}$ has a zero at $s=s_0$, and examine the behavior of the amplitude around $s=s_0$. Writing $\tilde{T}_{\text{tree}}=\cond{\sigma}^2T_{\text{tree}}$, the full amplitude can be expressed as
\begin{equation}
    T(s;x)
    = \frac{\tilde{T}_{\text{tree}}(s;x)}
    {\cond{\sigma}^2+G(s)\tilde{T}_{\text{tree}}(s;x)}  
    \label{eq:Ttilde} .
\end{equation}
Since we just factorize $\cond{\sigma}$, $\tilde{T}_{\text{tree}}$ also vanishes at $s=s_0$. Considering a small but finite $\cond{\sigma}$, the full amplitude $T(s;x)$ is always zero at $s=s_0$. For the energy $s=s_0+\epsilon$ with $|\epsilon|\sim \cond{\sigma}^2$, the denominator of the amplitude~\eqref{eq:Ttilde} can be expanded as
\begin{align}
    & \cond{\sigma}^2+\epsilon
    [G^{\prime}(s_0)
    \tilde{T}_{\text{tree}}(s_0;x)
    +G(s_0)
    \tilde{T}_{\text{tree}}^{\prime}(s_0;x)]
    +\mathcal{O}(\epsilon^2) \nonumber \\
    = & \cond{\sigma}^2+\epsilon G(s_0)
    \tilde{T}_{\text{tree}}^{\prime}(s_0;x)
    +\mathcal{O}(\epsilon^2)
    \label{eq:s0pole} , 
\end{align}
where $X^{\prime}(s_0)=\partial X(s)/\partial s|_{s=s_0}$ and we have used $\tilde{T}_{\text{tree}}(s_0;x)=0$. Eq.~\eqref{eq:s0pole} indicates that  the amplitude~\eqref{eq:Ttilde} has a pole at $\epsilon =-\cond{\sigma}^2/[G(s_0)\tilde{T}_{\text{tree}}^{\prime}(s_0;x)]$ which is complex since $G(s)$ is complex above the threshold. This is the pole toward which the higher energy pole moves in Fig.~\ref{fig:modelD}. Note also that the residue of this pole is $\epsilon\tilde{T}_{\text{tree}}^{\prime}(s_0;x)+\mathcal{O}(\epsilon^2)$ so the coupling to the $\pi\pi$ state also gradually vanishes. Therefore, when it approaches the real axis, this pole is not physically relevant, in the sense that it does not affect the spectrum very much. In the limit of $\cond{\sigma}\to 0$, $T(s;x)\to [G(s)]^{-1}$ and it is regular at $s=s_0$.

One may consider that this pole should move to the energy of the bare pole. Actually, in Ref.~\cite{Yokokawa:2002pw}, the softening phenomena of a similar model (denoted by ``\textit{Model B}'') is studied in the chiral limit. There are two poles in vacuum, and one of them moves to the origin, while the other moves to the bare pole when the symmetry is restored.

There is a difference in the treatment of the coupling constant; in Ref.~\cite{Yokokawa:2002pw}, the coupling constant of the $\sigma$ pole to the scattering state is kept fixed as $g_{\sigma}$, while the corresponding coupling in our model is proportional to $1/\cond{\sigma}^2$ as seen in the second term of Eq.~\eqref{eq:modelext}, so it varies with the symmetry restoration. For the expression~\eqref{eq:modelext}, the coupling $g_{\sigma}$ is proportional to $x/\cond{\sigma}$, so we can remove the $\cond{\sigma}$-dependence of the coupling constant by the replacement $x\to x\cond{\sigma}/\cond{\sigma}_0$. We checked that in this case the higher energy pole moves to the position of the bare state, namely, our model is qualitatively consistent with the analysis in Ref.~\cite{Yokokawa:2002pw} studied in the chiral limit. 

In summary, we have examined the softening of the $\sigma$ meson in four different models. By comparing model A and model B, we find that the softening of the dynamically generated $\sigma$ meson is qualitatively different from the $\sigma$ meson as the chiral partner. The formation of virtual state provides a novel softening phenomena for the dynamically generated sigma meson. In the energy region close to the threshold, the results in models C and D are similar to the model B, which can be understood by the dominance of the leading order term of low energy expansion. These observations leads to the following conclusions:
\begin{itemize}
\item If there is a bare $\sigma$ pole term which approaches to the $\pi\pi$ threshold energy region as we decrease the chiral condensate, the threshold enhancement is driven by the bare pole contribution and the spectral change and the pole trajectory will be those in model A (Fig.~\ref{fig:modelA}).
\item If the bare $\sigma$ pole does not exist in the threshold energy region even with the symmetry restoration (case II for the symmetry restoration), the near threshold spectrum is dominated by the dynamically generated $\sigma$ state and the results will be similar to those in model B (Fig.~\ref{fig:modelB}). The appearance of the virtual state is essential for the novel softening pattern.
\end{itemize}

\section{Summary}\label{sec:summary}

We study the properties of the $\sigma$ meson in the $\pi\pi$ scattering associated with the restoration of chiral symmetry, in order to extract the quantity which reflects the origin of the resonance. We show that, with the explicit symmetry breaking, the pattern of the threshold enhancement of the dynamically generated $\sigma$ meson is qualitatively different from the softening of the $\sigma$ meson as the chiral partner of the pion. The special nature of the $s$-wave resonance plays an essential role for this difference; as the symmetry is restored, the dynamically generated $\sigma$ resonance becomes a virtual state with a finite width, so the strong enhancement of the spectral function does not take place when the real part of the pole position of the $\sigma$ crosses the threshold, in contrast to the case with the chiral partner $\sigma$. When the virtual state turns into the bound state, the spectral function shows a sharp peak at the threshold, which takes place at a later stage of the symmetry restoration than the $\sigma$ meson as the chiral partner. 

We also consider several models with the CDD pole contribution which is driven by the QCD dynamics, such as the four quark state and the glueball. Analyzing these cases, we find that the softening pattern of the models, in which the mass of the bare $\sigma$ is unchanged with the symmetry restoration, is boiled down to the result of the dynamically generated $\sigma$. This is caused by the dominance of the leading order interaction of low energy expansion around the threshold. Since the low energy behavior of the amplitude is governed by the dynamics of the Nambu-Goldstone boson, this conclusion seems to be universal as long as the bare $\sigma$ pole stays in sufficiently higher energy region than the threshold.

We also study the mechanism of the dynamical generation of the resonance without symmetry restoration, decomposing the interaction kernel into the pole term and the contact term. The pole term always generates a resonance in the full amplitude through the coupling to the $\pi\pi$ scattering state, like the Feshbach resonance. It is shown that the contact term should contain some attractive component, in order to generate a resonance dynamically in addition to the state driven by the pole term of the interaction.

The property of the dynamically generated $\sigma$ meson in the symmetry restoration limit is investigated for the discussion of the chiral partner. We find that the mass of the dynamically generated $\sigma$ meson is degenerated with the pion in the restoration limit, and that the coupling strength to the $\pi\pi$ scattering state turns out to be proportional to $m_{\pi}$, which vanishes in the chiral limit $m_{\pi}\to 0$. The mass degeneracy with the pion and the vanishing of the coupling constant are the same behavior with the chiral partner of the pion in the symmetry restoration limit. It is rather nontrivial result that the dynamically generated $\sigma$ behaves like the chiral partner in the symmetry restoration limit.

The present framework grasps the essential feature of the $\pi\pi$ scattering, such as chiral symmetry, analyticity, and unitarity. We point out the important role of the finite pion mass for the property of the $\sigma$ meson associated with the chiral symmetry restoration. The finite pion mass changes the qualitative property of the softening, and the behavior of the $\sigma$ pole in the restoration limit. We should however keep in mind that there are some more effects to be included in the realistic situation, \textit{i.e.}, the $\rho$ meson in $t$-channel exchange, $\bar{K}K$ channel in three flavors, U$_A$(1) anomaly effect, and so on. Among others, it should be interesting to study the softening phenomena in a framework with exact crossing symmetry~\cite{Caprini:2008fc,Yndurain:2007qm,Caprini:2005zr}, since the effect of the crossed channels would be important for the softening behavior around the threshold and the bound state below the threshold. We hope that the present analysis provides a first step for the systematic study in more realistic framework.

\section*{Acknowledgments}

The authors are grateful to M. Oka and Y. Kanada-En'yo for useful discussion. T.H. thanks the support from the Global Center of Excellence Program by MEXT, Japan through the Nanoscience and Quantum Physics Project of the Tokyo Institute of Technology. This work was partly supported by the Grant-in-Aid for Scientific Research from MEXT and JSPS (Nos. 22740161, 22105507, 21840026, 20028004 and 20540273), and the Grant-in-Aid for the Global COE Program ``The Next Generation of Physics, Spun from Universality and Emergence'' from MEXT of Japan. This work was done in part under the Yukawa International Program for Quark-hadron Sciences (YIPQS).


\end{document}